\documentclass[11pt]{article}
\usepackage{amsmath,amssymb,epsfig,sint}

\newcommand{\R}{\rm I\kern-.2emR}
\newcommand{\C}{\rm \kern.25em\vrule height1.4ex
depth-.12ex width.06em\kern-.31em C}
\newcommand{\N}{{\rm I\kern-.16em N}}
\newcommand{\Z}{{\rm Z\kern-.35em Z}}

\newcommand{\rme}{{\rm e}}
\newcommand{\rmd}{{\rm d}}
\newcommand{\rmO}{{\rm O}}

\newcommand{\be}{\begin{equation}}   
\newcommand{\ex}{\end{equation}}
\newcommand{\ba}{\begin{eqnarray}}
\newcommand{\ea}{\end{eqnarray}}

\newcounter{subequation}[equation]
\makeatletter

\expandafter\let\expandafter
\reset@font\csname reset@font\endcsname

\def\subeqnarray{\arraycolsep1pt
    \def\@eqnnum\stepcounter##1{\stepcounter{subequation}%
        {\reset@font\rm(\theequation\alph{subequation})}}
\jot5mm     \eqnarray}

\makeatother
\newcommand{\msbar}{{\rm \overline{MS\kern-0.14em}\kern0.14em}}

\newcommand{\On}{O($n$)\ }
\newcommand{\lan}{$1/n$\ expansion\,\,}

\begin{document}
\begin{titlepage}

\begin{flushright}
MPP-2006-125\\
   January 2007
\end{flushright}

\vskip 0.20 true cm

%\vskip 0.20 true cm
\begin{center}
{\Large\bf 
Construction and clustering properties\\ of the 2-d non--linear 
$\sigma$--model form factors:\\ O$(3)$, O$(4)$, large $n$ examples} 
\end{center}
\vskip 1 true cm
\centerline{\large Janos Balog}
\vskip1ex
\centerline{Research Institute for Particle and Nuclear Physics}
\centerline{1525 Budapest 114, Pf. 49, Hungary}
\vskip 1 true cm
\centerline{\large Peter Weisz}
\vskip1ex
\centerline{Max-Planck-Institut f\"ur Physik}
\centerline{F\"ohringer Ring 6, D-80805 M\"unchen, Germany}
\vskip 1 true cm
\centerline{\bf Abstract}
\vskip 1.0ex
Multi--particle form factors of local operators in integrable models 
in two dimensions seem to have the property that they 
factorize when one subset of the particles in the external states are 
boosted by a large rapidity with respect to the others. 
This remarkable property,
which goes under the name of form factor clustering, was first observed 
by Smirnov in the O(3) non--linear $\sigma$--model and has  
subsequently found useful applications in integrable models 
without internal symmetry structure. In this paper we conjecture the 
nature of form factor clustering for the general \On $\sigma$--model 
and make some tests in leading orders of the  $1/n$ expansion
and for the special cases $n=3,4$.

\vfill
%\noindent{---------------}\\
\eject

\end{titlepage}

\section{Introduction}

In this paper we will investigate certain properties of 
form factors of integrable models in two dimensions.
This class of models allow a unique field theoretical insight 
because they admit special non--perturbative methods to their
solution known as the S--matrix bootstrap approach \cite{Smirnov,KW}. 
In particular once the spectrum of stable (massive)
states has been identified a well motivated S--matrix can be postulated.
Going further one can then attempt to solve equations for the
form factors of local operators and finally compute correlation
functions of these operators by saturating with a complete set 
of intermediate states. There are many examples of such applications. 
Of particular interest are the uncovering of structural relations which 
may have corresponding validity or inspire similar relations 
in models in higher dimensions.

We will in this paper consider mainly the non--linear \On 
$\sigma$--models which have the additional interesting feature
that they are asymptotically free. In particular for the case $n=3$
many form factors are explicitly known and one can compute 
vacuum 2--point functions up to rather high energies and compare
the results with numerical MC and analytic perturbative results \cite{BN}.

One of our main motivations for the present paper arose from our recent 
work on structure functions in these models \cite{BW}. A result that
particularly intrigues us concerns the small Feynman $x$ behavior ($Q^2$ 
fixed); we found that this was of the form $f(x)A(Q^2)$ where the behavior
of $f(x)$ at small $x$ reflects the high energy behavior of the
scattering amplitudes and the function $A(Q^2)$ is determined in 
terms of the vacuum 2--point function. We speculated that this
structure may be universal in asymptotically free field theories, 
in particular for QCD
\footnote{It holds for example in approximations like naive vector
meson dominance.}.   
One of the key properties in the derivation of the result (for the
$\sigma$--model) is the property of clustering of form factors,
and it may be that at least this property has some analogy in QCD.

Roughly, form factor clustering says that if one considers a 
multi--particle form factor of a local operator and one boosts a subset
of particles A uniformly with respect to the rest B by a large 
rapidity $\triangle$, then to leading order the form factor factorizes
into a function of $\triangle$ times a product of two functions, 
one depending only on the rapidities of A and the other only on the 
rapidities of B. 
The functions appearing here are again themselves form factors. 

To our knowledge the first observation of FF clustering was by
Smirnov \cite{Smirnov} for the O(3) non--linear $\sigma$--model. 
Thereafter
investigations of this structure were mainly pursued for S--matrices 
without internal symmetry structure (see e.g. \cite{KM}--\cite{AMV}).
Recently it was used in the construction of local operators 
in the Sinh--Gordon model by Delfino and Niccoli \cite{DN}. 
FF clustering appears
to be an extra constraint on form factors which can be
imposed in addition to the usual axioms. Its main application
so far has been to identify the operator associated with
a particular solution of the FF equations i.e. clustering can be 
useful in model building.

A systematic study of clustering properties for 
models with internal symmetry has not appeared in the literature so far. 
It is the purpose of this paper to make steps to fill this gap
for the case of the \On non--linear $\sigma$--models. 
In order to be able to make non-trivial tests of the clustering properties 
of the form factors we first had to work out some $\sigma$-model form factors
explicitly. In particular while considering the test in O(4) we had to 
work out details of the 3--particle spin form factor and this is presented in 
Appendix~D. While the structure (a tensor product of
SU(2) form factors) was given previously by Smirnov \cite{Smi2},
he only considered the case of form factors with an even number of
particles. We also work out a number of form factors in leading orders of
the large $n$ expansion both by applying bootstrap techniques for this
case and by the standard saddle point expansion of the functional integral.
A by-product of our investigation is the verification of the (expected, but 
non-trivial) equivalence of the two methods.
 
The organization of the paper is as follows. In the next section 
we give a brief introduction to the model, in particular the 
S--matrix is described and form factors of some familiar operators
are defined. In Sect.~3 we remind the reader of the FF axioms,
give the 2--particle form factors of  operators introduced in Sect.~2 and
consider the specific case of the 3--particle spin form factor.
In Sect.~4, for comparison, we first briefly review FF clustering 
for models without internal symmetry. 
Thereafter we consider the structure of FF clustering in the \On 
non--linear $\sigma$--models. We motivate an ansatz for the general 
form for various cases encountered, and conjecture a 
relation of the leading FF clustering behavior to the 
anomalous dimensions of the operators which is reminiscent to the 
form of operator product expansions valid at short distances.
In Sect.~5 we consider solutions for the form factors in leading orders
of the $1/n$ expansion, and in Sect.~6 we verify that these solutions are
indeed identical with results derived from the quantum field theoretic 
formalism. Tests of the ansatz in the 
leading order of the $1/n$--expansion are described in Sect.~7
and tests in the particular cases $n=3,4$ in Sect.~8. 
Various technical details are relegated to appendices. 

%\vfill
%\eject

\section{\On non--linear $\sigma$--model S--matrix and operators}

\subsection{The Zamolodchikov S--matrix}

Particles in the \On model are characterized by their mass $M$ and
the quantum numbers $(a,\theta)$, where $a=1,2,\dots,n$ is an \On 
vector index and $\theta$ is the rapidity of the particle in terms of 
which the components of its momentum are $p^0=M\cosh\theta$ and
$p^1=M\sinh\theta$. When two particles scatter there is no particle 
production and the bootstrap S--matrix proposed by
Zamolodchikov and Zamolodchikov \cite{ZZ} is of the form 
\begin{equation}
S_{ab}^{cd}(\theta)=\sigma_1(\theta)\,\delta^{cd}\,\delta_{ab}+
\sigma_2(\theta)\,\delta^c_a\,\delta^d_b+
\sigma_3(\theta)\,\delta^d_a\,\delta^c_b\,,
\label{Smat}
\end{equation}
where
\begin{equation} 
\sigma_1(\theta)=\frac{-2\pi i\chi}{i\pi-\theta}\,\sigma_2(\theta)\,,
\qquad\qquad
\sigma_3(\theta)=\frac{-2\pi i\chi}{\theta}\,\sigma_2(\theta)
\label{sigma13}
\end{equation}
and
\begin{equation}
\sigma_2(\theta)=\frac{-\theta}{\theta-2\pi i\chi}\,
\exp\left\{i\delta(\theta)\right\}\,,
\label{sigma2}
\end{equation}
where the phase appearing here is given by
\begin{equation}
\delta(\theta)=2\int_0^\infty\,\frac{\rmd\omega}{\omega}\,
\sin(\theta\omega)\,\tilde K_n(\omega)\,.
\label{delta}
\end{equation}
with kernel
\begin{equation} 
\tilde K_n(\omega)=
\frac{\rme^{-\pi\omega}+\rme^{-2\pi\chi\omega}}
{1+\rme^{-\pi\omega}}\,.
\label{tildeK}
\end{equation}
We have used the notation $\chi=\frac{1}{n-2}$ in the above formulae.

It is useful to introduce the invariant  amplitudes 
corresponding to $s$--channel ``isospin'' $I=0,2,1$:
\begin{eqnarray}
S_0(\theta)&=&n\sigma_1(\theta)+\sigma_2(\theta)+\sigma_3(\theta)\,,\nonumber\\
S_2(\theta)&=&\sigma_2(\theta)+\sigma_3(\theta)=
-\exp\{i\delta(\theta)\}\,,\label{S012}\\
S_1(\theta)&=&-\sigma_2(\theta)+\sigma_3(\theta)\,,\nonumber\\
\nonumber
\end{eqnarray}
which obey unitarity $S_I(\theta)S_I(-\theta)=1$.
%We have
%\begin{equation}
%S_1(\theta)=\frac{2\pi i+(n-2)\theta}{2\pi i-(n-2)\theta}\,S_2(\theta),
%\qquad\qquad S_0(\theta)=\frac{i\pi+\theta}{i\pi-\theta}\,S_1(\theta).
%\end{equation}

The particular cases $n=3,4$ are discussed in more detail in Appendix~A.

\subsection{Operators and form factors}

In this subsection we discuss the form factors of the most important
operators in the model, those of the \On spin field, the Noether
current and the energy--momentum tensor. We also define the form
factors of a symmetric, traceless scalar operator.

\subsubsection{The \On field}

The conventional normalization of the \On field is given by its
one--particle matrix elements:
\begin{equation}
\langle0\vert\Phi^a(0)\vert b,\theta\rangle=\delta^{ab}\,.
\end{equation}
The general $r$--particle matrix elements define its form factors by
\begin{equation}
\langle0\vert\Phi^a(0)\vert
b_1,\theta_1;\dots;b_r,\theta_r\rangle^{{\rm in}}=
\Lambda_n\,f^a_{b_1\dots b_r}(\theta_1,\dots,\theta_r)\,,
\end{equation}
where
\begin{equation}
\Lambda_3=\frac{2}{\sqrt{\pi}},\qquad\qquad \Lambda_n=1,\quad n>3\,.
\end{equation}
The physical \lq\lq in" states correspond to the rapidity ordering
$\theta_1>\theta_2>\dots>\theta_r$. The form factors are originally
defined for this ordered set of real rapidities but can be extended to
the complete complex (multi)--rapidity space by analytic continuation.
See Sect.~\ref{axioms}. We use the state normalization
\begin{equation}
%\begin{split}
\phantom{\rangle}^{{\rm in}}\langle a^\prime_1,\theta^\prime_1;\dots;
a^\prime_r,\theta^\prime_r\vert
a_1,\theta_1;\dots;a_r,\theta_r\rangle^{{\rm in}}
=(4\pi)^r\delta_{a^\prime_1a_1}\dots\delta_{a^\prime_ra_r}
\delta(\theta^\prime_1-\theta_1)\dots\delta(\theta^\prime_r-\theta_r)\,.
%\end{split}
\end{equation}

\subsubsection{The Noether current}

The normalization of the Noether current operators $J^{ab}_\mu(x)$ is 
fixed by the equal time commutation relations
\begin{equation}
\left[J^{ab}_0(0,x),\Phi^c(0,y)\right]=it^{ab}_{cd}\,
\delta(x-y)\,\Phi^d(0,y)\,,
\label{CR}
\end{equation}
where
\be
t^{cd}_{ab}=\delta^c_a\,\delta^d_b-\delta^d_a\,\delta^c_b\,.
\end{equation}
The current form factors are given by
\begin{equation}
\langle0\vert J^{ab}_\mu(0)\vert b_1,\theta_1;\dots;b_r,\theta_r
\rangle^{{\rm in}}=
-i\epsilon_{\mu\alpha}q^\alpha\,
f^{ab}_{b_1\dots b_r}(\theta_1,\dots,\theta_r)\,,
\label{currFF}
\end{equation}
where
\begin{equation}
q^\alpha=\left(p_1+p_2+\cdots+p_r\right)^\alpha,\qquad
p_i=(p_i^0,p_i^1)=(M\cosh\theta_i,M\sinh\theta_i)
\end{equation}
and $\epsilon_{01}=-\epsilon_{10}=1$.
The normalization (\ref{CR}) implies\footnote{Recall that for particles
with rapidity $\theta$ corresponding to \lq bra' vectors in the 
expectation value the form factor functions have to be analytically
continued to the complex rapidity value $\theta+i\pi$.}
the following result for the one--particle expectation value.
\begin{equation}
\langle c,\theta\vert J^{ab}_\mu(0)\vert d,\theta\rangle=
-2ip_\mu t^{ab}_{cd}\,.
\label{expect}
\end{equation}

\subsubsection{The energy--momentum tensor}

The energy--momentum tensor is normalized so that its space integral
\begin{equation}
H=\int_{-\infty}^\infty\,{\rm d}x\,T_{00}(0,x)
\end{equation}
is the Hamiltonian of the system with one--particle eigenvalues given by
$H\vert b,\theta\rangle=M\cosh\theta\vert b,\theta\rangle$.
The energy--momentum tensor form factors are
\begin{equation}
\langle0\vert T_{\mu\nu}(0)\vert b_1,\theta_1;\dots;b_r,\theta_r
\rangle^{{\rm in}}=
(\eta_{\mu\nu}q^2-q_\mu q_\nu)\,f_{b_1\dots b_r}(\theta_1,\dots,\theta_r)\,,
\end{equation}
where $\eta_{\mu\nu}$ is the 1+1 dimensional metric characterized by 
$\eta_{00}=-\eta_{11}=1$.

The case $n=3$ is discussed in further detail in Appendix~B.

\subsubsection{Symmetric, traceless tensor operator}

Finally we define the form factors of a Lorenz scalar and symmetric,
traceless iso--tensor operator $\Sigma^{cd}$
\begin{equation}
\langle0\vert\Sigma^{cd}(0)\vert b_1,\theta_1;\dots;b_r,\theta_r
\rangle^{{\rm in}}=
\tilde f^{cd}_{b_1\dots b_r}(\theta_1,\dots,\theta_r)\,.
\label{tensorFF}
\end{equation}

\subsection{Two--particle form factors}

Using \On symmetry and Poincar\'e invariance, the two--particle form
factors can be parameterized as follows
\begin{eqnarray}
\langle0\vert J^{cd}_\mu(0)\vert a,\alpha;b,\beta\rangle&=&
i\epsilon_{\mu\nu}\,q^\nu\,\psi_1(\alpha-\beta)\,t^{cd}_{ab}\,,\nonumber\\
\langle0\vert\Sigma^{cd}(0)\vert a,\alpha;b,\beta\rangle&=&
-i\psi_2(\alpha-\beta)\,\tilde t^{cd}_{ab}\,,\label{2ff}\\
\langle0\vert T_{\mu\nu}(0)\vert a,\alpha;b,\beta\rangle&=&
\frac{i}{2}\,(q_\mu q_\nu-q^2\eta_{\mu\nu})\,
\psi_0(\alpha-\beta)\,\delta_{ab}\,,\nonumber\\
\nonumber
\end{eqnarray}
where
\begin{equation}
s^{cd}_{ab}=\delta^c_a\,\delta^d_b+\delta^d_a\,\delta^c_b,\qquad
\tilde t^{cd}_{ab}=s^{cd}_{ab}-\frac{2}{n}\,\delta^{cd}\,\delta_{ab}\,.
\end{equation}
It can be shown that the normalization of the operators defined
above implies the following singularity structure for the functions
$\psi_i(\theta)$
\begin{eqnarray}
\psi_0(\theta)&\approx&\frac{-4i}{(\theta-i\pi)^2}\,,\qquad\theta\approx
i\pi,\nonumber\label{norm}\\
\psi_1(\theta)&\approx&\frac{2}{\theta-i\pi}\,,\qquad
\ \ \ \, \theta\approx
i\pi\,,\\
\psi_2(\theta)&&{\rm regular\ at\ \ \ }\,\quad\theta=i\pi\,.\nonumber\\
\nonumber
\end{eqnarray}

%\vfill
%\eject

\section{Form factor axioms}
\label{axioms}

In this section we recall the functional equations \cite{Smirnov}
satisfied by the scalarized form factors, which we generically denote by
${\cal F}_{a_1\dots a_r}(\theta_1,\dots,\theta_r)$ in this section.
It turns out to be convenient to introduce the Faddeev--Zamolodchikov
operators $Z^+_a(\theta)$ satisfying the exchange relation
\begin{equation} 
Z^+_a(\theta)Z^+_b(\theta^\prime)=S_{ab}^{yx}(\theta-\theta^\prime)
Z^+_x(\theta^\prime)Z^+_y(\theta)\,.
\end{equation} 
Now we can define the multi--index matrix
$S_{ba_1\dots a_r;b_1\dots b_ra}(\beta\vert\theta_1,\dots,\theta_r)$
by the relation
\begin{equation} 
Z^+_b(\beta)Z^+_{a_1}(\theta_1)\cdots Z^+_{a_r}(\theta_r)
=S_{ba_1\dots a_r;b_1\dots b_ra}(\beta\vert\theta_1,\dots,\theta_r)
Z^+_{b_1}(\theta_1)\cdots Z^+_{b_r}(\theta_r)Z^+_a(\beta)\,.
\end{equation} 

The form factor axioms are the following five functional equations 
\cite{Smirnov}
\begin{equation} 
{\cal F}_{a_1\dots a_r}(\theta_1,\dots,\theta_r)=
{\cal F}_{a_1\dots a_r}(\theta_1+\lambda,\dots,\theta_r+\lambda)\,,
\label{AX1}
\end{equation} 
\begin{equation} 
{\cal F}_{\cdots xy\cdots}(\cdots\theta,\theta^\prime\cdots)=
S_{xy}^{vw}(\theta-\theta^\prime)
{\cal F}_{\cdots wv\cdots}(\cdots\theta^\prime,\theta\cdots)\,,
\label{AX2}
\end{equation} 
\begin{equation} 
{\cal F}_{a_1a_2\dots a_r}(\theta_1+2\pi i,\theta_2,\dots,\theta_r)=
{\cal F}_{a_2\dots a_ra_1}(\theta_2,\dots,\theta_r,\theta_1)\,,
\label{AX3}
\end{equation} 
\begin{equation} 
\begin{split}
&\lim_{\varepsilon\to0}\varepsilon\,{\cal F}_{aba_1\dots a_r}
(\beta+i\pi+\varepsilon,\beta,\theta_1,\dots,\theta_r)\\
&=2i\left\{
\delta_{ab}{\cal F}_{a_1\dots a_r}(\theta_1,\dots,\theta_r)-
S_{ba_1\dots a_r;b_1\dots b_ra}(\beta\vert\theta_1,\dots,\theta_r)
{\cal F}_{b_1\dots b_r}(\theta_1,\dots,\theta_r)\right\}\,,
\end{split}
\label{AX4}
\end{equation} 
\begin{equation} 
{\cal F}_{a_1\dots a_r}(\theta_1,\dots,\theta_r)=w_p
{\cal F}_{a_r\dots a_1}(-\theta_r,\dots,-\theta_1)\,.
\label{AX5}
\end{equation} 
In the last equation $w_p$ is the parity of the scalarized form
factors. It is equal to unity for all operators considered above 
except for the Noether current, for which it is equal to $-1$.

Next we define a new type of reduced form factors 
\footnote{Here ``new" is wrt those usually defined by factoring out the
product of 2--particle scalar form factors e.g as in Appendix~B} by
\ba 
{\cal F}_{a_1\dots a_r}(\theta_1,\dots,\theta_r)&=&
\frac{F_{a_1\dots a_r}(\theta_1,\dots,\theta_r)}
{C_r(\theta_1,\dots,\theta_r)}\,,
\label{reducedF}
\\
C_r(\theta_1,\dots,\theta_r)&\equiv&
\prod_{1\le i<j\le r}\cosh\left(\frac{\theta_i-\theta_j}{2}\right)\,.
\ea
Three of the form factor equations for 
$F_{a_1\dots a_r}(\theta_1,\dots,\theta_r)$
are of the same form as (\ref{AX1}), (\ref{AX2}) and (\ref{AX5}) and
the equation corresponding to (\ref{AX3}) is only modified by a sign factor
$(-1)^{r-1}$. Finally the residue axiom (\ref{AX4}) is rewritten as
\begin{equation} 
\begin{split}
&F_{aba_1\dots a_r}
(\beta+i\pi,\beta,\theta_1,\dots,\theta_r)=\left(\frac{i}{2}\right)^r\,
\prod_{j=1}^r\sinh(\beta-\theta_j)\cdot\\
&\left\{S_{ba_1\dots a_r;b_1\dots b_ra}(\beta\vert\theta_1,\dots,\theta_r)
F_{b_1\dots b_r}(\theta_1,\dots,\theta_r)
-\delta_{ab}F_{a_1\dots a_r}(\theta_1,\dots,\theta_r)\right\}\,.\\
\end{split}
\label{ax4}
\end{equation} 

\subsection{Two--particle form factors}

Two of the form factor equations, (\ref{AX1}) and (\ref{AX5}), are
automatically satisfied by the ansatz (\ref{2ff}). The residue
equation (\ref{AX4}) does not apply to two--particle form factors, but
one has to satisfy the normalization conditions (\ref{norm}) instead.
Finally (\ref{AX2}) and (\ref{AX3}) become ($I=0,1,2$)
\begin{equation}
\psi_I(\theta)=S_I(\theta)\,\psi_I(-\theta)
\end{equation}
and
\begin{equation}
\psi_I(\theta+2\pi i)=(-1)^I\,\psi_I(-\theta)
\end{equation}
respectively where the $S_I$ are defined in (\ref{S012}).

A useful building block in the construction of form factors is the function
\begin{equation}
\Delta(\theta)=\int_0^\infty\,\frac{{\rm d}\omega}{\omega}\,
\tilde K_n(\omega)\,\frac{\cosh[(\pi+i\theta)\omega]-1}{\sinh\pi\omega}\,.
\label{Delta}
\end{equation}
Its main properties are
\begin{equation}
\Delta(i\pi+\theta)=\Delta(i\pi-\theta),\qquad \Delta(i\pi)=0,\qquad
\Delta(\theta)=\Delta(-\theta)+i\delta(\theta)
\end{equation}
and its asymptotic behavior for large positive $\theta$ is given by
\begin{equation}
\Delta(i\pi+\theta),\,{\mathcal Re}\Delta(\theta) \approx -\frac{\theta}{2}
\tilde K_n(0)-\frac{\ln\theta}{\pi}\tilde K^\prime_n(0)+{\rm O}(1)\,.
\end{equation}
If in (\ref{delta}) we substitute $\tilde K_n(\omega)$ by the function 
$\tilde k_\alpha(\omega)=-{\rm e}^{-\pi\alpha\omega}$ we get 
\begin{equation}
{\rm e}^{i\delta_\alpha(\theta)}=\frac{i\alpha\pi+\theta}{i\alpha\pi-\theta}\,.
\end{equation}
We denote the related building block by $\Delta_\alpha(\theta)$.

Using the building blocks defined above we find
\begin{eqnarray}
\psi_1(\theta)&=&\tanh\frac{\theta}{2}\,\exp\{\Delta(\theta)
+\Delta_{2\chi}(\theta)\}\approx c_1\,\theta^{-\chi}\,,\nonumber\\
\psi_2(\theta)&=&\sinh\frac{\theta}{2}\,\exp\{\Delta(\theta)\}
\approx c_2\,\theta^{\chi}\,,\label{psi120}\\
\psi_0(\theta)&=&\frac{-2i}{i\pi-\theta}\,\psi_1(\theta)\approx c_0\,
\theta^{-(1+\chi)}\,.\nonumber\\
\nonumber
\end{eqnarray}
Here we also indicate the asymptotic behavior of the form factors 
$\psi_I(\theta)$ for large positive $\theta$.

\subsection{Three--particle form factors}
\label{3FF}

In this section we write down the form factor equations for the
three--particle form factors of the \On field operators.
We start with
the definition 
\begin{equation}
f^d_{abc}(\alpha,\beta,\gamma)=\frac{F^d_{abc}(\alpha,\beta,\gamma)}
{\cosh\left(\frac{\alpha-\beta}{2}\right)\,
\cosh\left(\frac{\alpha-\gamma}{2}\right)\,
\cosh\left(\frac{\beta-\gamma}{2}\right)}\,.
\end{equation}
The form factor equations (\ref{AX1}--\ref{AX5}) in this special case
become ($n>3$) 
\begin{eqnarray}
F^d_{abc}(\alpha,\beta,\gamma)&=&
F^d_{abc}(\alpha+\lambda,\beta+\lambda,\gamma+\lambda)\,,\label{3ff1}\\
F^d_{abc}(\alpha,\beta,\gamma)&=&
S^{yx}_{bc}(\beta-\gamma)\,
F^d_{axy}(\alpha,\gamma,\beta)\,,\label{3ff2}\\
F^d_{abc}(\alpha,\beta,\gamma)&=&
F^d_{bca}(\beta,\gamma,\alpha-2\pi i)\,,\label{3ff3}\\
F^d_{abc}(\beta+i\pi,\beta,\gamma)&=&
\frac{i}{2}\sinh(\beta-\gamma)\,\left\{S^{ad}_{bc}(\beta-\gamma)
-\delta_{ab}\delta^d_c\right\}\,,\label{3ff4}\\
F^d_{abc}(\alpha,\beta,\gamma)&=&
F^d_{cba}(-\gamma,-\beta,-\alpha)\,.\label{3ff5}\\
\nonumber
\end{eqnarray}
(\ref{3ff1}) and (\ref{3ff3}) are satisfied by the following ansatz
\begin{equation}
\begin{split}
F^d_{abc}&(\alpha,\beta,\gamma)=
\delta_{ad}\delta_{bc}X(\alpha-\beta,\alpha-\gamma)\\
&+\delta_{bd}\delta_{ac}X(\beta-\gamma,\beta-\alpha+2\pi i)+
\delta_{cd}\delta_{ab}X(\gamma-\alpha+2\pi i,\gamma-\beta+2\pi i)\,.\\
\label{Xdef}
\end{split}
\end{equation}
We can also rewrite (\ref{3ff4}) and (\ref{3ff5}) in terms of this
single function $X$. We get
\begin{eqnarray}
X(u,v)&=&X(2\pi i-v,2\pi i-u)\,,\nonumber\\
X(\theta,i\pi)&=&\frac{i}{2}\sinh\theta\,\sigma_3(\theta),\label{3ff45}\\
X(\theta,i\pi+\theta)&=&\frac{i}{2}\sinh\theta\,
\left[\sigma_2(\theta)-1\right]\,.
\nonumber\\
\nonumber
\end{eqnarray}
Finally (\ref{3ff2}) is equivalent to
\begin{equation}
\begin{split}
X(u,v)&=\left[n\sigma_1(v-u)+\sigma_2(v-u)+\sigma_3(v-u)\right]\,X(v,u)\\
&+\sigma_1(v-u)\,\left[X(u-v,2\pi i-v)+X(2\pi i-u,2\pi i+v-u)\right]\\
\end{split}
\label{sX1}
\end{equation}
and
\begin{equation}
\begin{split}
X(v-u,2\pi i-u)&=
\sigma_2(v-u)\,X(2\pi i-u,2\pi i+v-u)\\
&+\sigma_3(v-u)\,X(u-v,2\pi i-v)\,.\\
\end{split}
\label{sX2}
\end{equation}

%\vfill
%\eject

\section{Form factor clustering}

\subsection{Models without internal symmetry}

We first briefly review clustering properties of form factors 
for models without internal symmetry for which the majority
of detailed investigations have been carried out so far.  
In \cite{DSC} Delfino, Simonetti and Cardy studied models defined as
perturbations of conformal invariant theories i.e.
those formally defined by the action
\be
A=A_{\rm CFT}+g\int\rmd^2x\,\phi(x)
\end{equation}
where the operator $\phi(x)$ is of dimension $2\delta<2$. Further 
attention is restricted to those perturbations where an infinite number
of integrals of motion survive and the resulting massive model
is integrable. 

Consider first the case when there is only one species of massive 
particle of mass $m$ and 2--particle S--matrix element $S(\theta)$.
The {\it cluster hypothesis} proposes that multi--particle 
form factors of a scaling operator $\Phi$ of dimension $2\delta_\Phi<2$ 
in such a theory factorize according to
\be
\lim_{\Lambda\to\infty} F^\Phi_{r+l}
(\theta_1+\Lambda,\dots,
\theta_r+\Lambda,\theta_{r+1},\dots,\theta_{r+l})
=\frac{1}{\langle\Phi\rangle}
F^\Phi_{r}(\theta_1,\dots,\theta_r)
F^\Phi_{l}(\theta_{r+1},\dots,\theta_{r+l})
\label{Fact0}
\end{equation}
Actually, (\ref{Fact0}) is only valid for operators corresponding to 
primaries in the conformal limit. 
More generally, conformal operators can be classified
as
\begin{equation}
{\cal L}_m\,\bar{\cal L}_{\bar m}\Phi,
\end{equation}
where ${\cal L}_m$ is a combination (at level $m$) of the left Virasoro
operators and similarly $\bar{\cal L}_{\bar m}$ is built from right
Virasoro operators. Identifying operators in the massive theory with
their conformal limit, the generalization of (\ref{Fact0}) for descendant
operators reads \cite{DN1}
\begin{equation}
\begin{split}
\lim_{\Delta\to\infty} {\rm e}^{-m\Delta} F_{r+l}^{{\cal L}_m\bar{\cal L}_m
\Phi}(\theta_1&+\Delta,\dots,\theta_r+\Delta,\theta_{r+1},\dots,
\theta_{r+l})\\
&=\frac{1}{\langle\Phi\rangle}\,
F_{r}^{{\cal L}_m \Phi}(\theta_1,\dots,\theta_r)\,
F_{l}^{\bar{\cal L}_m \Phi}(\theta_{r+1},\dots,\theta_{r+l})
\end{split}
\end{equation}
It follows that knowledge of the multi--particle FF can be used to 
determine the vacuum expectation value $\langle\Phi\rangle$ 
which is in general non--vanishing due to the absence of internal 
symmetries. This is useful since the vev can be
obtained by other means e.g. by the thermodynamic Bethe ansatz.

Ref.~\cite{DSC} made the validity of (\ref{Fact0}) highly plausible
by considering the massless limit to the UV critical point in
which the mass $m\to0$ and rapidities of the particles are
simultaneously taken to $\infty$ such that momenta are fixed.
The basic well--accepted assumptions are that 
i) $\lim_{\theta\to\infty}S(\theta)=1$ so that massless left and right 
movers decouple and ii)
that the operator space of the conformal point and that of the perturbed
theory have the same basic structure. In particular to the scaling 
operator $\Phi(x)$ in the off--critical theory there is an associated  
conformal operator $\tilde{\Phi}$ of the same scaling 
dimension $2\delta_\Phi$. The property has been noticed in the past 
to be fulfilled by various FF solutions in specific models 
\cite{KM}--\cite{DN}, 
and many examples and tests of FF clustering have been successfully performed.

Tests of the hypothesis for a given S--matrix, involve solving the general
functional equations together with the cluster constraints and seeing 
whether the number of independent solutions equals the number in the 
corresponding Kac table of the associated CFT. Once this has been 
established one can identify the operators corresponding to the
solutions by computing the dimensions $\delta_\Phi$ either by 
studying the short distance behavior of the 2--point function 
computed by saturation by lowest states, or using the DSC \cite{DSC} sum 
rule
\be
\delta_\Phi^{UV}-\delta_\Phi^{IR}
=-\frac{1}{4\pi\delta}\int\rmd^2x\langle\Theta(x)\Phi(0)\rangle_c\,,
\label{TWi}
\end{equation}
where $\Theta(x)$ is the trace of the energy momentum tensor which
is related to the perturbing field by $\Theta=4\pi g(1-\delta)\phi(x)$.
In a massive theory $\delta_\Phi^{IR}=0$.

\subsection{FF clustering in the non--linear sigma model}

To our knowledge FF clustering in the O(3) non--linear sigma
model was first discussed by Smirnov \cite{Smirnov}. 
A more detailed exposition
of this case was presented by Balog and Niedermaier \cite{BN}. 
In the following we consider the general \On case which exhibits 
a rather rich structure.

\subsection{Clustering (leading term)}

Let us divide the particles into two subsets and boost the particles
in the first set by a large (positive) rapidity $\Delta$. Form factor 
clustering means that the scalarized, dimensionless form factors have
universal large $\Delta$ asymptotics (which usually behave as a power
in $\Delta$ instead of the constant behavior exhibited by the models
considered in the previous subsection):
\begin{equation}
\begin{split}
&f_{a_1\dots a_kb_1\dots b_l}
(\alpha_1+\Delta,\dots,\alpha_k+\Delta,
\beta_1,\dots,\beta_l)\\
&\qquad\cong h_{k;l}(\Delta)\,
g_{a_1\dots a_k;b_1\dots b_l}(\alpha_1,\dots,\alpha_k;
\beta_1,\dots,\beta_l) 
%\qquad({\rm no\ sum\ in\ }A),
\end{split}
\label{clusterW}
\end{equation}
where the clustering function 
\begin{equation}
h_{k;l}(\Delta+\lambda)\cong h_{k;l}(\Delta)
\end{equation}
and also the functional form (and the dependence on $\Delta$) of the
sub--leading terms depend on the type of the operator. These sub--leading 
terms are often suppressed by some negative power of $\Delta$, see
Sect. \ref{cluCORR}. 

Using the asymptotic properties of the S--matrix, 
\be
S_{cd}^{ab}(\theta)\cong \delta^a_c \delta^b_d +
\frac{2\pi i\chi}{\theta}\,t^{ac}_{bd}+\dots,
\label{Sasy}
\end{equation}
we can show that the expressions
\begin{equation}
g_{a_1\dots a_k;b_1\dots b_l}(\alpha_1,\dots,\alpha_k;
\beta_1,\dots,\beta_l) 
\end{equation}
satisfy the form factor axioms (\ref{AX1})--(\ref{AX4}) in the
variables corresponding to the first set while the dependence on the
particles belonging to the second set are playing the role of 
dummy parameters.
This is almost trivial for (\ref{AX1})--(\ref{AX3}), and for (\ref{AX4}) 
we get
\begin{equation}
\begin{split}
&f_{aba_1\dots a_kb_1\dots b_l}(\gamma+i\pi+\epsilon+\Delta,\gamma+\Delta,
\alpha_1+\Delta,\dots,\alpha_k+\Delta,\beta_1,\dots,\beta_l)\\
&\qquad\cong h_{k+2;l}(\Delta)\,g_{aba_1\dots a_k;b_1\dots b_l}
(\gamma+i\pi+\epsilon,\gamma,
\alpha_1,\dots,\alpha_k;\beta_1,\dots,\beta_l)\\
&\qquad\cong \frac{2i}{\epsilon}\big\{\delta_{ab}
\,f_{a_1\dots a_kb_1\dots b_l}
(\alpha_1+\Delta,\dots,\alpha_k+\Delta,\beta_1,\dots,\beta_l)\\
&\qquad\qquad
-S_{ba_1\dots a_kb_1\dots b_l;\tilde a_1\dots \tilde a_k\tilde
b_1\dots\tilde b_la}(\gamma+\Delta\vert
\alpha_1+\Delta,\dots,\alpha_k+\Delta,\beta_1,\dots,\beta_l)\\
&\qquad\qquad\qquad
\,\cdot f_{\tilde a_1\dots \tilde a_k\tilde b_1\dots \tilde b_l}
(\alpha_1+\Delta,\dots,\alpha_k+\Delta,\beta_1,\dots,\beta_l)\big\}\\
&\qquad\cong \frac{2i}{\epsilon}\,h_{k;l}(\Delta)\big\{\delta_{ab}
\,g_{a_1\dots a_k;b_1\dots b_l}
(\alpha_1,\dots,\alpha_k;\beta_1,\dots,\beta_l)\\
&\qquad\qquad
-S_{ba_1\dots a_k;\tilde a_1\dots \tilde a_ka}(\gamma\vert
\alpha_1,\dots,\alpha_k)
\,g_{\tilde a_1\dots \tilde a_k;b_1\dots b_l}
(\alpha_1,\dots,\alpha_k;\beta_1,\dots,\beta_l)\big\}\,.
\end{split}
\end{equation}
From this we see that, because of the uniqueness of the solution of  
the set of form factor axioms,
\begin{equation}
h_{k+2;l}(\Delta)=h_{k;l}(\Delta)
\label{recu}
\end{equation}
and for $k\geq1$
\begin{equation}
\begin{split}
&g_{aba_1\dots a_k;b_1\dots b_l}
(\gamma+i\pi+\epsilon,\gamma,
\alpha_1,\dots,\alpha_k,\beta_1,\dots,\beta_l)\\
&\qquad\cong \frac{2i}{\epsilon}\,\big\{\delta_{ab}
\,g_{a_1\dots a_k;b_1\dots b_l}
(\alpha_1,\dots,\alpha_k;\beta_1,\dots,\beta_l)\\
&\qquad\qquad
-S_{ba_1\dots a_k;\tilde a_1\dots \tilde a_ka}
(\gamma\vert\alpha_1,\dots,\alpha_k)
\,g_{\tilde a_1\dots \tilde a_k;b_1\dots b_l}
(\alpha_1,\dots,\alpha_k;\beta_1,\dots,\beta_l)\big\}\,,
\end{split}
\end{equation}
which is nothing but the residue axiom for the first set of particles
where the quantum numbers of the particles belonging to the second set
are dummy parameters.

Similarly we find that
\begin{equation}
h_{k;l+2}(\Delta)=h_{k;l}(\Delta)
\label{recu2}
\end{equation}
and the axioms (\ref{AX1})--(\ref{AX4}) are satisfied in the
variables corresponding to the second set, for fixed
$\{a_1\dots a_k\}$, $\{\alpha_1,\dots,\alpha_k\}$.

Using the recursion relations (\ref{recu}) and (\ref{recu2}) we see
that there are three clustering families, the cases of odd--odd, 
even--odd and even--even clustering.

The function $g_{a_1\dots a_kb_1\dots b_l}$ is essentially
a product of two scalarized form factors, corresponding to the
operators $B$ and $C$. Denoting the original operator by $A$, 
the clustering relations can be symbolically represented as
\be
A\sim h(\Delta)\,\, B \bullet C\,,
\label{symb}
\end{equation}
or, since in the \On model $h(\Delta)$ is always a power, more
explicitly as 
\be
A\sim \frac{1}{\Delta^\kappa}\,\, B \bullet C\,.
\label{symb2}
\end{equation}
The residue axiom is applicable for clustering for $k\geq3$ only but
from \cite{BW} we see that for \On nonsinglets
\begin{equation}
h_{2;l}(\Delta)=\frac{1}{\Delta}
\end{equation}
since
\begin{equation}
F^A_{ab;b_1\dots b_l}(\Delta+i\pi+\epsilon,\Delta,\beta_1,\dots,\beta_l)
\cong \frac{4\pi\chi}{\epsilon\Delta}\,\,t^{ab}_{AB}\,\,
F^B_{b_1\dots b_l}(\beta_1,\dots,\beta_l)\,.
\label{nons}
\end{equation}
where $t^{ab}_{AB}$ are \On generators in the representation under 
consideration. Similarly 
\begin{equation}
h_{k;2}(\Delta)=\frac{1}{\Delta}
\end{equation}
and
\begin{equation}
g^A_{a_1\dots a_k;ab}(\alpha_1,\dots,\alpha_k;\gamma+i\pi+\epsilon,\gamma)
\cong -\frac{4\pi\chi}{\epsilon}\,t^{ab}_{AB}\,
f^B_{a_1\dots a_k}(\alpha_1,\dots,\alpha_k)\,.
\end{equation}

For $k$ odd, using the uniqueness of the solution, we can solve the
problem step by step, starting from the $k=1$ case:
\begin{equation}
g_{a;b_1\dots b_l}(\alpha;\beta_1,\dots,\beta_l)=
G_{ab_1\dots b_l}(\beta_1,\dots,\beta_l)\,.
\end{equation}
Then
\begin{equation}
g_{a_1\dots a_k;b_1\dots b_l}(\alpha_1,\dots,\alpha_k;
\beta_1,\dots,\beta_l)=
f^a_{a_1\dots a_k}(\alpha_1,\dots,\alpha_k)\,
G_{ab_1\dots b_l}(\beta_1,\dots,\beta_l)\,,
\end{equation}
where $f^a_{a_1\dots a_k}(\alpha_1,\dots,\alpha_k)$ is the form
factor of the basic field $\Phi^a$. 

Similarly for $l$ odd
\begin{equation}
g_{a_1\dots a_k;b_1\dots b_l}(\alpha_1,\dots,\alpha_k;
\beta_1,\dots,\beta_l)=
H_{a_1\dots a_kb}(\alpha_1,\dots,\alpha_k)\,
f^b_{b_1\dots b_l}(\beta_1,\dots,\beta_l)\,,
\end{equation}
where
\begin{equation}
g_{a_1\dots a_k;b}(\alpha_1,\dots,\alpha_k;\beta)=
H_{a_1\dots a_kb}(\alpha_1,\dots,\alpha_k)\,.
\end{equation}

\subsection{Odd--odd clustering}
\label{odd-odd}

The simplest case is the odd--odd clustering. Starting from
\begin{equation}
g_{a;b}(\alpha;\beta)=T_{ab}
\end{equation}
we can build
\begin{equation}
g_{a_1\dots a_k;b_1\dots
b_l}(\alpha_1,\dots,\alpha_k;\beta_1,\dots,\beta_l)
=T_{ab}\,f^a_{a_1\dots a_k}(\alpha_1,\dots,\alpha_k)
\,f^b_{b_1\dots b_l}(\beta_1,\dots,\beta_l)\,.
\end{equation}

In the case of the current operator $J^{cd}$ 
\begin{equation}
h(\Delta)=c_1\,\Delta^{-\chi}
\end{equation}
and
\begin{equation}
T_{ab}\qquad\rightarrow\qquad -t^{cd}_{ab}\,.
\end{equation}
Similarly for the symmetric tensor $\Sigma^{cd}$
\begin{equation}
h(\Delta)=c_2\,\Delta^{\chi}
\end{equation}
and
\begin{equation}
T_{ab}\qquad\rightarrow\qquad -i\tilde t^{cd}_{ab}\,.
\end{equation}
Finally for the energy--momentum tensor $T$ we have
\begin{equation}
h(\Delta)=c_0\,\Delta^{-(1+\chi)}
\end{equation}
and
\begin{equation}
T_{ab}\qquad\rightarrow\qquad -\frac{i}{2}\,\delta_{ab}\,.
\end{equation}

We can represent the above clustering relations symbolically as
\be
\begin{split}
J^{ab}&\quad\sim\quad\frac{1}{\Delta^\chi}\,
t^{ab}_{cd}\,\Phi^c\bullet \Phi^d\,,\\
\Sigma^{ab}&\quad\sim\quad\Delta^\chi\,
\tilde{t}^{ab}_{cd}\,\Phi^c\bullet \Phi^d\,,\\
T&\quad\sim\quad\frac{1}{\Delta^{1+\chi}}\,\Phi^a\bullet \Phi^a\,.
\end{split}
\end{equation}

\subsection{$3\,\rightarrow\, 2+1$ clustering}

This is the simplest case of even--odd clustering. In the odd--odd case
we could afford the luxury of using the exact solution of the
2--particle form factors to obtain the $2\,\rightarrow\, 1+1$ 
clustering relations. The exact 3--particle form factor is not known for
general $n$ (except for $n=3,4$ and in the large $n$ limit), but
the form factor equations can be solved in the clustering limit.
We start from the representation
\be
f^x_{abc}(\Delta,\beta,\gamma)=
\left(\Delta-\frac{\beta+\gamma}{2}\right)^{-\kappa}
\left\{A^x_{abc}(\beta-\gamma)+
\frac{\tilde A^x_{abc}(\beta-\gamma)}{\Delta-\frac{\beta+\gamma}{2}}+\dots
\right\}\,.
\label{CLUansatz}
\end{equation}
The functional form of the ansatz (\ref{CLUansatz}) is motivated by
the following considerations. First of all, it is easy to show \cite{DeMu}
using the known short distance asymptotics of the 2--point function
and the spectral representation that the spin form factor must not
grow faster than any power of the momenta, i.e. it is smaller than
$\rme^{\epsilon\Delta}$ for any $\epsilon$. This observation,
the known behavior of the 2--particle form factors and
the fact that the theory is asymptotically free, taken together make
plausible that for large momenta also the 3--particle (and higher)
form factors vary logarithmically.

Now we impose the form factor axioms in the large $\Delta$ limit. 
This gives for the leading coefficient $A^x_{abc}$ the equations
\be
A^x_{abc}(2\pi i-\theta)=A^x_{acb}(\theta)
\label{ex1}
\end{equation}
and
\be
A^x_{abc}(\theta)=S^{vu}_{bc}(\theta)\,A^x_{auv}(-\theta)\,,
\label{ex2}
\end{equation}
which means that, as expected, $A^x_{abc}$ satisfies the 
2--particle form factor
equations in the last two variables. The general solution is
\be
A^x_{abc}(\theta)=k_0\,\psi_0(\theta)\,\delta^x_a\,\delta_{bc}+
k_1\,\psi_1(\theta)\,t^{xa}_{bc}+
k_2\,\psi_2(\theta)\,\tilde t^{xa}_{bc}\,,
\label{gensol}
\end{equation}
where the $k_I$ are constants. In the $\kappa=1$ case we also have to
satisfy the residue axiom
\be
A^x_{cab}(i\pi+\epsilon)\cong-\frac{4\pi\chi}{\epsilon}\,t^{ab}_{xc}\,,
\end{equation}
which gives $k_1=-2\pi\chi$.

At the next order we find that $\tilde A^x_{abc}$ also satisfies
(\ref{ex2}) but instead of (\ref{ex1}) we have in this case
\be
\tilde A^x_{abc}(2\pi i-\theta)-\tilde A^x_{acb}(\theta)=
-i\pi\kappa\,A^x_{acb}(\theta) +2\pi i\chi\,t^{va}_{ub}\,A^x_{vcu}(\theta)\,.
\label{ex3}
\end{equation}
Let us define
\be
K^x_{abc}(\theta)\equiv\tilde A^x_{abc}(2\pi i-\theta)
-\tilde A^x_{acb}(\theta)\,.
\end{equation}
Obviously,
\be
 K^x_{acb}(2\pi i-\theta)+K^x_{abc}(\theta)=0\,.
\end{equation}
There is no solution for $\tilde A^x_{abc}$ unless the right hand side
of (\ref{ex3}) is also antisymmetric under this operation.
Thus the consistency of the next-to-leading order gives an additional 
condition on the leading order: 
\be
\chi\,t^{vb}_{ua}\,A^x_{vuc}+\chi\,t^{vb}_{uc}\,A^x_{vau}-
\kappa\,A^x_{bac}=0\,.
\end{equation}
This purely algebraic relation restricts the possible solutions 
(\ref{gensol}) so that only one of the coefficients $k_I$ can be 
different from zero and fixes the exponent $\kappa$ as follows: 
\be
\begin{split}
I=0 \quad {\rm case} \quad (k_0\not=0):\qquad\quad &\kappa=0,\\
I=1 \quad {\rm case} \quad (k_1\not=0):\qquad\quad &\kappa=1,\\
I=2 \quad {\rm case} \quad (k_2\not=0):\qquad\quad &\kappa=1+2\chi\,.
\end{split}
\end{equation}
Furthermore, the residue conditions give additional restrictions.
First of all, since $\psi_0(\theta)$ has a double pole at $\theta=i\pi$,
this excludes the $k_0\not=0$ ($\kappa=0$) solution.
In the $I=1$ ($\kappa=1$) case the residue conditions fix $k_1$ 
as given above. Finally, since $\psi_2(\theta)$ is regular, the
coefficient $k_2$ cannot be determined by this method.

Putting everything together, we get for the
$3\,\rightarrow\, 2+1$ clustering the following result:
\be
f^x_{abc}(\Delta,\beta,\gamma)
\cong -\frac{2\pi\chi}{\Delta}\,\psi_1(\beta-\gamma)\,t^{xa}_{bc}
+i\chi H_2\left(\frac{2\pi}{\Delta}\right)^{1+2\chi}\,
\psi_2(\beta-\gamma)\,\tilde t^{xa}_{bc}+\dots
\label{321clu0}
\end{equation}
Here we have reparameterized the constant $k_2$ (in terms of the new 
constant $H_2$) for later convenience.

(\ref{321clu0}) can also be written in terms of the full 2--particle
form factors as
\be
f^x_{abc}(\Delta,\beta,\gamma)
\cong \frac{2\pi\chi}{\Delta}\,f^{xa}_{bc}(\beta,\gamma)
-\chi H_2\left(\frac{2\pi}{\Delta}\right)^{1+2\chi}\,
\tilde f^{xa}_{bc}(\beta,\gamma)+\dots
\label{321clu}
\end{equation}
 
Note that the second piece in the clustering formula
(\ref{321clu}) is always subleading to the first. Actually, it
makes sense only for $n>4$, since for $n=3,4$ it
is not dominant over the $1/\Delta^2$ correction corresponding to the
first term. On the other hand, the two terms are close for larger
$n$ values and they become degenerate in the large $n$
limit. In Sect.~7 we check (\ref{321clu}) in the large
$n$ expansion.

Now we calculate these $1/\Delta^2$ correction terms in the large rapidity 
expansion. We take $\kappa=1$ and get from (\ref{CLUansatz})
\begin{equation}
f^x_{abc}(\Delta,\beta,\gamma)\approx\frac{1}{\Delta-\frac{\beta+\gamma}{2}}
\left\{A^x_{abc}(\xi)
+\frac{1}{\Delta}\widetilde A^x_{abc}(\xi)+\cdots\right\}\,,
\end{equation}
where $\xi=\beta-\gamma$. The leading term is:
\begin{equation}
A^x_{abc}(\xi)=-2\pi\chi\psi_1(\xi)\,t^{xa}_{bc}\,.
\end{equation}
To calculate the next term we first write
\begin{equation}
\widetilde A^x_{abc}(\xi)=2i\pi^2\chi\psi_1(\xi)\left\{
\lambda_1(\xi)\delta^x_a\delta_{bc}+
\lambda_2(\xi)\delta^x_b\delta_{ac}+
\lambda_3(\xi)\delta^x_c\delta_{ab}\right\}\,.
\end{equation}
We also introduce
\begin{equation}
\lambda_\pm=\lambda_2\pm\lambda_3,\qquad\qquad
\overline\lambda=n\lambda_1+\lambda_+\,.
\end{equation}
We now rewrite (\ref{ex3}) and (\ref{ex2}) for $\widetilde A^x_{abc}$ in
terms of these variables and get
\begin{equation}
\begin{split}
\lambda_+(\xi)&=\frac{2\pi i\chi-\xi}{2\pi i\chi+\xi}\,\lambda_+(-\xi)\,,\\
\lambda_-(\xi)&=\lambda_-(-\xi),\\
\overline\lambda(\xi)&=\frac{i\pi+\xi}{i\pi-\xi}\,\overline\lambda(-\xi)\,,
\end{split}\qquad\qquad\quad
\begin{split}
\lambda_+(2\pi i-\xi)&+\lambda_+(\xi)=2\,,\\
\lambda_-(2\pi i-\xi)&-\lambda_-(\xi)=0,\\
\overline\lambda(2\pi i-\xi)&+\overline\lambda(\xi)=4(1+\chi)\,.
\end{split}
\label{homlam}
\end{equation}
From the residue equations we get
\begin{equation}
\lambda_+(i\pi)=1\,,\qquad
\lambda_-(i\pi)=-2,\qquad
\overline\lambda(i\pi)=2(1+\chi)\,.\qquad
\label{reslam}
\end{equation}
This results from expanding the residue equation
\begin{equation}
f^x_{abc}(\alpha,\beta,\Delta)\approx\frac{2i}{\xi-i\pi}\,\left\{
\delta_{ab}\delta^x_c-S^{ax}_{bc}(\beta-\Delta)\right\}
\end{equation}
for large $\Delta$ using the expansion
\begin{equation}
S^{xy}_{ab}(\Delta)=\delta^x_a\delta^y_b+
\frac{2\pi i\chi}{\Delta}\,t^{xa}_{yb}-
\frac{2\pi^2\chi^2}{\Delta^2}\,t^{ua}_{vb}\,t^{xu}_{yv}+\cdots
\end{equation}

It is easy to find the general solution of (\ref{homlam}) and (\ref{reslam}).
We get
\begin{equation}
\lambda_-(\xi)=\ell_-(\cosh\xi);\qquad\quad \ell_-(-1)=-2\,.
\end{equation}
Since we know that $\lambda_-(\xi)$ must not grow exponentially for large
$\xi$ and it should be regular we conclude that
\begin{equation}
\lambda_-(\xi)=-2\,.
\end{equation}
We also get
\begin{equation}
\overline\lambda(\xi)=\frac{\overline\ell(\cosh\xi)}{i\pi-\xi}-
\frac{i(1+\chi)}{\pi}\,(i\pi+\xi);\qquad\quad \overline\ell(-1)=0
\end{equation}
and requiring regularity here gives
\begin{equation}
\overline\lambda(\xi)=-\frac{i(1+\chi)}{\pi}\,(i\pi+\xi)\,.
\end{equation}
Similarly we can find the regular solution of the $\lambda_+$ equations:
\begin{equation}
\lambda_+(\xi)=i\frac{2\pi i\chi-\xi}{\pi(1-2\chi)}+\omega(\chi)\,
\cosh\frac{\xi}{2}\,{\rm e}^{-\Delta_{2\chi}(\xi)}\,,
\label{plus1}
\end{equation}
where regularity at infinity requires that $\omega(\chi)$ is constant.
(\ref{plus1}) is only valid for $n\not=4$, because for $n=4$ it becomes 
singular. For $n=4$ the solution of the $\lambda_+$ equations is
\begin{equation}
\lambda_+(\xi)=\frac{2i}{\pi}(i\pi-\xi)\left\{g_0
+\frac{1}{4}\Psi\left(\frac{1}{2}+\frac{\xi}{2\pi i}\right)
+\frac{1}{4}\Psi\left(\frac{1}{2}-\frac{\xi}{2\pi i}\right)\right\}\,,
\label{plus2}
\end{equation}
where, again, regularity requires that $g_0$ cannot depend on the
relative rapidity $\xi$. 

$\omega(\chi)$ can be calculated for $\chi=1$ and $\chi=0$
(corresponding to $n=3$ and $n=\infty$ respectively) from the known
solutions or for general $n\not=4$ from the consistency of the 
next-to-next-to-leading order equations in the large rapidity expansion.
Consistency would not fix the value of $g_0$ for $n=4$, and, as we
will see later, it is actually not a constant, but depends (linearly)
on $\log\Delta$. In the light of the $1/(n-4)$ singularity in 
(\ref{plus1}) the presence of the logarithmic term in (\ref{plus2}) 
is not surprising. For $n=4$ (and only in this case) the $1/\Delta^2$
subleading term is accompanied by a $\log(\Delta)/\Delta^2$ term. One can
see that this is consistent here, because the term containing $g_0$ is 
exactly the same as the one corresponding to the constant $H_2$, 
the coefficient of the $\kappa=1+2\chi$ term in (\ref{321clu0}). 
The only way to determine $g_0$ is to solve the full O$(4)$ form 
factor equations explicitly. We will consider this problem in 
Section~8 and Appendix~D.

\subsection{General clustering}
\label{cluCORR}

What we can learn from the three-particle example is that
it is useful to include in (\ref{clusterW}) some of the
subleading terms as well. Thus we have to allow for the occurrence of
several terms (labeled by an index $\rho$) of the form
\begin{equation}
\begin{split}
&f_{a_1\dots a_kb_1\dots b_l}
(\alpha_1+\Delta,\dots,\alpha_k+\Delta,
\beta_1,\dots,\beta_l)\\
&\qquad\cong \sum_\rho h^{(\rho)}_{k;l}(\Delta)\,
g^{(\rho)}_{a_1\dots a_k;b_1\dots b_l}(\alpha_1,\dots,\alpha_k;
\beta_1,\dots,\beta_l)\,, 
%\qquad({\rm no\ sum\ in\ }A)\,,
\end{split}
\label{clusterC}
\end{equation}
where, as before, the coefficient functions  
$g^{(\rho)}$ have to solve the form factor equations for both sets
of variable.

In the case of odd--odd clustering the sum contains only one term so
the results in Section~\ref{odd-odd} do not change. 
But the general even--odd clustering formula ($k$ odd, $l$ even)
consists of two terms:
\be
\begin{split}
f^x_{a_1\dots a_kb_1\dots b_l}&(\alpha_1+\Delta,\dots,\alpha_k+\Delta,
\beta_1,\dots,\beta_l)\\
&\cong \frac{2\pi\chi}{\Delta}\,f^u_{a_1\dots a_k}(\alpha_1,\dots,
\alpha_k)\,f^{xu}_{b_1\dots b_l}(\beta_1,\dots,\beta_l)\\
&\qquad -\chi H_2
\left(\frac{2\pi}{\Delta}\right)^{1+2\chi}\,
f^u_{a_1\dots a_k}(\alpha_1,\dots,
\alpha_k)\,\tilde f^{xu}_{b_1\dots b_l}(\beta_1,\dots,\beta_l)+\dots
\end{split}
\end{equation}
This can be symbolically represented as
\be
\Phi^x\sim\frac{1}{\Delta}\,\Phi^u\bullet J^{xu}\,\,+\,\,
\frac{1}{\Delta^{1+2\chi}}\,\Phi^u\bullet\Sigma^{xu}\,.
\end{equation}

\subsection{A conjecture}

We have seen that clustering relations can be represented in the form
\be
A\sim \frac{1}{\Delta^\kappa}\,B\bullet C
\end{equation}
or as a sum of similar terms on the right hand side.
We found that the value of the exponent $\kappa$  depends on the
anomalous dimensions of the operators involved. In all cases we
studied so far we have the relation
\be
\kappa=d_B+d_C-d_A\quad ({\rm mod\ }1)\,, 
\end{equation}
where
\be
d_O=\frac{\gamma_O}{2\beta_0}\,.
\end{equation}
Here $\gamma_O$ is the coefficient of the first term (in perturbation
theory) of the anomalous dimension of the operator $O$ and $\beta_0$
is the coefficient of the first term of the perturbative 
$\beta$--function.

Explicitly,
\be
\begin{split}
d_\Phi&=\frac{1}{2}\,\frac{\gamma_0}{2\beta_0}=\frac{1}{2}(1+\chi)\,,\\
d_J&=d_T=0,\\
d_\Sigma&=\frac{\gamma_\Sigma}{2\beta_0}=2\chi\,.
\end{split}
\end{equation}
Here we used the results
\be
\gamma_0=\frac{n-1}{2\pi},\qquad\quad
\beta_0=\frac{n-2}{4\pi},\qquad\quad
\gamma_\Sigma=\frac{1}{\pi}\,.
\end{equation}

Using this conjecture, it is possible to write down the formula 
for even--even clustering without doing any calculation. For the current
operator we  get
\be
\begin{split}
&f^{xy}_{a_1\dots a_kb_1\dots b_l}
(\alpha_1+\Delta,\dots,\alpha_k+\Delta,\beta_1,\dots,\beta_l)\\
&\cong \frac{2\pi\chi}{\Delta}\left[
f^{xq}_{a_1\dots a_k}(\alpha)
f^{yq}_{b_1\dots b_l}(\beta)
-f^{yq}_{a_1\dots a_k}(\alpha)
f^{xq}_{b_1\dots b_l}(\beta)\right]\\
&+\Omega\chi\left(\frac{2\pi}{\Delta}\right)^{1+4\chi}
\left[\tilde f^{xq}_{a_1\dots a_k}(\alpha)
\tilde f^{yq}_{b_1\dots b_l}(\beta)
-\tilde f^{yq}_{a_1\dots a_k}(\alpha)
\tilde f^{xq}_{b_1\dots b_l}(\beta)\right]\,.
\end{split}
\end{equation}
Here we have used (beyond \On symmetry) the fact that isospin 0 
form factors cannot occur here (they would give a double pole in the
residue axiom) and that the residue axiom fixes the coefficient
of the $1/\Delta$ term containing the current form factors. The
power $1+4\chi$ is consistent with the conjecture and is uniquely fixed
by the requirement that the two terms should be degenerate in the large
$n$ limit. The constant $\Omega$ is not fixed by these considerations,
but by studying the $k=l=2$ case in the large $n$ limit we can show that 
\be
\Omega=1+\rmO\left(\frac{1}{n}\right)\,.
\end{equation}
Analogously for the tensor form factor we get
\be
\begin{split}
&\tilde f^{xy}_{a_1\dots a_kb_1\dots b_l}
(\alpha_1+\Delta,\dots,\alpha_k+\Delta,\beta_1,\dots,\beta_l)\\
&\cong -\frac{2\pi\chi}{\Delta}\left[
f^{xq}_{a_1\dots a_k}(\alpha)
\tilde f^{yq}_{b_1\dots b_l}(\beta)
+f^{yq}_{a_1\dots a_k}(\alpha)
\tilde f^{xq}_{b_1\dots b_l}(\beta)\right]\\
&+\frac{2\pi\chi}{\Delta}\left[
\tilde f^{xq}_{a_1\dots a_k}(\alpha)
f^{yq}_{b_1\dots b_l}(\beta)
+\tilde f^{yq}_{a_1\dots a_k}(\alpha)
f^{xq}_{b_1\dots b_l}(\beta)\right]\\
&+\tilde\Omega\chi\left(\frac{2\pi}{\Delta}\right)^{1+2\chi}
\Big[\tilde f^{xq}_{a_1\dots a_k}(\alpha)
\tilde f^{yq}_{b_1\dots b_l}(\beta)
+\tilde f^{yq}_{a_1\dots a_k}(\alpha)
\tilde f^{xq}_{b_1\dots b_l}(\beta)\\
&\qquad\qquad -\frac{2}{n}\delta^{xy}
\tilde f^{pq}_{a_1\dots a_k}(\alpha)
\tilde f^{pq}_{b_1\dots b_l}(\beta)\Big]\,,
\end{split}
\end{equation}
where
\be
\tilde\Omega=-2+\rmO\left(\frac{1}{n}\right)\,.
\end{equation}

%\vfill
%\eject

\section{Bootstrap form factors in leading orders \lan}

In this section we consider the solution of the form factor
equations in leading order \lan. We start with the $1/n$ expansion of the
S--matrix and 2--particle form factors.

\subsection{S--matrix and \lan}

The \lan\ of the coefficients in (\ref{Smat}) is of the form
\begin{equation}
\sigma_i(\theta)=\delta_{i2}+\frac{a_i(\theta)}{n}
+\frac{b_i(\theta)}{n^2}+\rmO\left(\frac{1}
{n^3}\right),\qquad\quad i=1,2,3
\end{equation}
with
\begin{equation}
a_1(\theta)=-\frac{2\pi i}{i\pi-\theta},\qquad\qquad
a_2(\theta)=-\frac{2\pi i}{\sinh\theta},\qquad\qquad
a_3(\theta)=-\frac{2\pi i}{\theta},\label{a123}
\end{equation}
\begin{equation}
b_1(\theta)=\frac{-4\pi}{i\pi-\theta}\left(i+\frac{\pi}{\sinh\theta}\right)\,,
\qquad\qquad
b_3(\theta)=\frac{-4\pi}{\theta}\left(i+\frac{\pi}{\sinh\theta}\right)\,,
\end{equation}
and ($\Psi(z)=\Gamma^\prime(z)/\Gamma(z)$)
\begin{equation}
\begin{split}
b_2(\theta)=&-\frac{4\pi i}{\sinh\theta}
-\frac{2\pi^2}{\sinh^2\theta}\\
&+\frac{1}{2}\left[
\Psi^\prime\left(\frac{1}{2}+\frac{i\theta}{2\pi}\right)
-\Psi^\prime\left(\frac{1}{2}-\frac{i\theta}{2\pi}\right)
-\Psi^\prime\left(1+\frac{i\theta}{2\pi}\right)
+\Psi^\prime\left(-\frac{i\theta}{2\pi}\right)\right]\,.
\end{split}
\end{equation}
Later we will need the combination
\begin{equation}
b_1(\theta)+a_2(\theta)+a_3(\theta)=\frac{\theta+i\pi}{\theta-i\pi}
\left(\frac{2\pi i}{\theta}-\frac{2\pi i}{\sinh\theta}\right)\,.
\end{equation}

\subsection{Large $n$ expansion of the two--particle form factors}

The large $n$ expansion of the two--particle form factors is given by
the expansion of the functions $\psi_I$ in (\ref{psi120}):
\begin{eqnarray}
\psi_0(\theta)&=& \frac{2i\tanh\frac{\theta}{2}}{i\pi-\theta}\,\left\{
1+\frac{2\pi}{n}\left[a(\theta)+b(\theta)\right]+{\rm O}
\left(\frac{1}{n^2}\right)\right\}\,,\nonumber\\
\psi_1(\theta)&=& \tanh\frac{\theta}{2}\,\left\{
1+\frac{2\pi}{n}\left[a(\theta)+b(\theta)\right]+{\rm O}
\left(\frac{1}{n^2}\right)\right\}\,,\label{psiasy}\\
\psi_2(\theta)&=& i\,\left\{
1+\frac{2\pi}{n}\left[a(\theta)-b(\theta)\right]+{\rm O}
\left(\frac{1}{n^2}\right)\right\}\,,\nonumber\\
\nonumber
\end{eqnarray}
where
\begin{equation}
a(\theta)=\frac{1}{2\pi}+\frac{\theta-i\pi}{2\pi\sinh\theta}
\end{equation}
and
\begin{equation}
b(\theta)=\frac{i}{2\theta}-\frac{1}{4\pi}\left[
\Psi\left(\frac{i\theta}{2\pi}\right)+
\Psi\left(-\frac{i\theta}{2\pi}\right)-2\Psi\left(\frac{1}{2}\right)\right]\,.
\end{equation}

\subsection{Large $n$ expansion of the spin 3--particle form factor}
\label{3FFn}

We assume that for large $n$ the function $X$ appearing in (\ref{Xdef})
has an expansion of the form:
\begin{equation}
X(u,v)=\frac{f(u,v)}{n}+\frac{g(u,v)}{n^2}+\rmO\left(\frac{1}{n^3}\right)\,.
\end{equation}
The form factor equations (\ref{3ff45}) have to be satisfied order by
order in the expansion. On the other hand, (\ref{sX1}) and (\ref{sX2})
mix the expansion coefficients (beyond leading order). At leading
order they lead to
\begin{eqnarray}
f(u,v)&=&\left[1+a_1(v-u)\right]\,f(v,u)\,,\\
f(u,v)&=&f(v,2\pi i+u)\\
\nonumber
\end{eqnarray}
and at next-to-leading order we have
\begin{equation}
\begin{split}
g(u,v)&=\left[1+a_1(v-u)\right]\,g(v,u)\\
&+\left[b_1(v-u)+a_2(v-u)+a_3(v-u)\right]\,f(v,u)\\
&+a_1(v-u)\left[f(u-v,2\pi i-v)+f(2\pi i-u,2\pi i+v-u)\right]
\end{split}
\label{nlo1}
\end{equation}
and
\begin{equation}
\begin{split}
&g(v-u,2\pi i-u)=g(2\pi i-u,2\pi i+v-u)\\
&+a_2(v-u)\,f(2\pi i-u,2\pi i+v-u)
+a_3(v-u)\,f(u-v,2\pi i-v)\,.\\
\end{split}
\label{nlo2}
\end{equation}

\subsubsection{Leading order solution}

We here summarize the equations the leading order form factor $f(u,v)$
has to satisfy.
\begin{eqnarray}
f(u,v)&=&f(2\pi i-v,2\pi i-u)\,,\label{LO1}\\
f(\theta,i\pi)&=&\frac{i}{2}\,\sinh\theta\,a_3(\theta)=
\frac{\pi\sinh\theta}{\theta}\,,\\
f(\theta,i\pi+\theta)&=&\frac{i}{2}\,\sinh\theta\,a_2(\theta)=\pi,\\
f(u,v)&=&\frac{v-u+i\pi}{v-u-i\pi}\,f(v,u)\,,\\
f(u,v)&=&f(v,2\pi i+u)\,.\label{LO5}\\
\nonumber
\end{eqnarray}

Defining the function
\begin{equation}
R(\theta)\equiv\frac{\pi\sinh\theta}{i\pi -\theta}\,,
\end{equation}
with properties
\begin{equation}
R(i\pi-\theta)=\frac{\pi\sinh\theta}{\theta},\qquad\quad 
R(i\pi)=\pi,\qquad\quad
R(2\pi i-\theta)=R(\theta)\,,
\end{equation}
we now take the ansatz
\begin{equation}
f(u,v)=R(v-u)\,\left[s(u,v)+1\right]
\end{equation}
and verify that (\ref{LO1}--\ref{LO5}) require
\begin{equation}
\begin{split}
s(u,v)=s(v,u)&=s(2\pi i+u,v)=s(-u,-v),\\
s(\theta,i\pi)&=s(\theta,i\pi+\theta)=0\,.\\
\end{split}
\label{zero}
\end{equation}
It is easy to see that regularity and boundedness at infinity
allows the trivial solution $s(u,v)=0$ only leading to the unique 
leading order solution
\begin{equation}
f(u,v)=R(v-u)\,.
\label{LO}
\end{equation}

\subsubsection{Next-to-leading order solution}

Using the leading order solution (\ref{LO}) the form factor equations
(\ref{nlo1}) and (\ref{nlo2}) can be simplified a little. We list here
the complete set of next-to-leading order (NLO) form factor equations
after this simplification.
\begin{eqnarray}
g(u,v)&=&\left[1+a_1(v-u)\right]\,g(v,u)\nonumber\\
&+&\left[b_1(v-u)+a_2(v-u)+a_3(v-u)\right]\,R(u-v)\label{a}\\
&+&a_1(v-u)\,\left[R(u)+R(v)\right]\,,\nonumber\\
g(u,v)&=&g(v,2\pi i+u)+a_2(u)R(v-u)+a_3(u)R(v),\label{b}\\
g(u,v)&=&g(2\pi i-v,2\pi i-u)\,,\label{c}\\
g(\theta,i\pi)&=&\frac{i}{2}\sinh\theta\,b_3(\theta),\label{d}\\
g(\theta,i\pi+\theta)&=&\frac{i}{2}\sinh\theta\,b_2(\theta)\,.\label{e}\\
\nonumber
\end{eqnarray}

We present the NLO solution in several steps in order to make the
checking of equations (\ref{a}--\ref{e}) easier. We start with
\begin{equation}
\begin{split}
g(u,v)&=\left\{G(u,v)+\frac{i\pi}{v-u}-\frac{i\pi}{\sinh(v-u)}-
\frac{\pi}{R(u)}-\frac{\pi}{R(v)}\right\}\,R(v-u)\\
&-R(u)-R(v)\,.\\
\end{split}
\end{equation}
It is easy to show that (\ref{a}) is satisfied if $G$ satisfies
\begin{equation}
G(u,v)=G(v,u)\,.
\end{equation}
Next we write
\begin{equation}
\begin{split}
G&(u,v)=S(u,v)+\frac{v-u}{\sinh(v-u)}-k(u-v)-k(v-u)\\
&+\frac{\sinh v}{\sinh(v-u)}\left\{
\frac{u-i\pi}{i\pi-v}+2k(u)-k(v+i\pi)-k(v-i\pi)\right\}\\
&-\frac{\sinh u}{\sinh(v-u)}\left\{
\frac{v-i\pi}{i\pi-u}+2k(v)-k(u+i\pi)-k(u-i\pi)\right\}\,.\\
\end{split}
\end{equation}
Here 
\begin{equation}
k(\theta)=\frac{1}{2}\,\Psi\left(-\frac{i\theta}{2\pi}\right)\,.
\end{equation}
(\ref{a}) and (\ref{b}) are satisfied if
\begin{equation}
S(u,v)=S(v,u)=S(u+2\pi i,v)\,.
\end{equation}
In the next step we write
\begin{equation}
\begin{split}
S(u,v)&=\Sigma(u,v)+\frac{i\pi}{2\sinh(v-u)}\Big\{
\cosh u-\cosh v+\frac{\sinh v(1+\cosh u)}{\sinh u}\\
&-\frac{\sinh u(1+\cosh v)}{\sinh v}\Big\}\,.\\
\end{split}
\end{equation}
In addition to (\ref{a}) and (\ref{b}), (\ref{c}) is also satisfied if
\begin{equation}
\Sigma(u,v)=\Sigma(v,u)=\Sigma(u+2\pi i,v)=\Sigma(-u,-v)\,.
\end{equation}
Finally in the last step we represent $\Sigma$ as
\begin{equation}
\Sigma(u,v)=\frac{\cosh u+\cosh v}{1+\cosh(u-v)}+3+
\Psi\left(\frac{1}{2}\right)+\sigma(u,v)\,.
\end{equation}
It now follows that all equations (\ref{a}-\ref{e}) are satisfied
if $\sigma(u,v)$ satisfies the same equations as $s(u,v)$ in (\ref{zero}).

Although the form factors are bounded and regular functions for all $n$, 
the large $n$ expansion, as can be seen from (\ref{a123}), introduces 
some singularities at rapidity differences equal to $0$ or $i\pi$.
Nevertheless, one can show that (\ref{zero}) has only trivial solution
for $s(u,v)$ even if one allows (first order) poles at these special
rapidity differences. Thus $\sigma(u,v)=0$ and the NLO solution is unique.

\subsection{$n=\infty$ form factor equations}

In this section we write down the form factor equations in the
leading order of the large $n$ expansion. In this limit the
homogeneous equations take the form
\begin{equation} 
F_{a_1\dots a_r}(\theta_1,\dots,\theta_r)=
F_{a_1\dots a_r}(\theta_1+\lambda,\dots,\theta_r+\lambda)\,,
\label{NAX1}
\end{equation} 
\begin{equation} 
F_{\cdots xy\cdots}(\cdots\theta,\theta^\prime\cdots)=
\frac{1}{n}a_1(\theta-\theta^\prime)
F_{\cdots zz\cdots}(\cdots\theta^\prime,\theta\cdots)\delta_{xy}+
F_{\cdots yx\cdots}(\cdots\theta^\prime,\theta\cdots)\,,
\label{NAX2}
\end{equation} 
\begin{equation} 
F_{a_1a_2\dots a_r}(\theta_1+2\pi i,\theta_2,\dots,\theta_r)=
(-1)^{r-1}\,F_{a_2\dots a_ra_1}(\theta_2,\dots,\theta_r,\theta_1)\,,
\label{NAX3}
\end{equation} 
\begin{equation} 
F_{a_1\dots a_r}(\theta_1,\dots,\theta_r)=w_p
F_{a_r\dots a_1}(-\theta_r,\dots,-\theta_1)\,.
\label{NAX5}
\end{equation} 
In (\ref{NAX2}) the first term on the right hand side is of O$(1)$ if
the contracted indices belong to the same Kronecker delta. Otherwise
it is of order $1/n$ and can be dropped.

To calculate the residue equation to leading order we first note the
recursion relation
\begin{equation} 
S_{ba_1\dots a_r;b_1\dots b_ra}(\beta\vert\theta_1,\dots,\theta_r)=
S_{ba_1\dots a_{r-1};b_1\dots b_{r-1}x}(\beta\vert\theta_1,\dots,\theta_{r-1})
\,S_{xa_r}^{ab_r}(\beta-\theta_r)\,.
\end{equation} 
Starting from the $r=1$ case
\begin{equation} 
\begin{split}
S_{ba_1;b_1a}(\beta\vert\theta_1)&=\delta_{ab}\delta_{a_1b_1}+
\frac{1}{n}a_1(\beta-\theta_1)\delta_{ab_1}\delta_{ba_1}\\
&+\frac{1}{n}a_2(\beta-\theta_1)\delta_{ab}\delta_{a_1b_1}+
\frac{1}{n}a_3(\beta-\theta_1)\delta_{aa_1}\delta_{bb_1}
+{\rm O}\left(\frac{1}{n^2}\right)
\end{split}
\end{equation} 
it is easy to show by induction that the residue equation
takes the form
\begin{equation} 
\begin{split}
nF_{aba_1\dots a_r}&
(\beta+i\pi,\beta,\theta_1,\dots,\theta_r)\\
=\left(\frac{i}{2}\right)^r\Big\{\prod_{j=1}^r&\sinh(\beta-\theta_j)
\Big\}\Bigg[\Big\{\sum_{k=1}^ra_2(\beta-\theta_k)\Big\}\,
\delta_{ab}F_{a_1\dots a_r}(\theta_1,\dots,\theta_r)\\
&+\sum_{k=1}^ra_1(\beta-\theta_k)\,\delta_{ba_k}\,
F_{a_1\dots a\dots a_r}^{\ \ \ (k)}
(\theta_1,\dots,\theta_k,\dots,\theta_r)\\
&+\sum_{k=1}^ra_3(\beta-\theta_k)\,\delta_{aa_k}\,
F_{a_1\dots b\dots a_r}^{\ \ \ (k)}
(\theta_1,\dots,\theta_k,\dots,\theta_r)\\
+\frac{1}{n}\sum_{l<k}a_1(\beta-\theta_l)&a_3(\beta-\theta_k)
\,\delta_{aa_k}\delta_{ba_l}\,
F_{a_1\dots z\dots z\dots a_r}^{\ \ \ \,(l)\,(k)}
(\theta_1,\dots,\theta_l,\dots,\theta_k,\dots,\theta_r)\Bigg]
\end{split}
\label{NAX4}
\end{equation} 
in the leading order of the large $n$ expansion. Again, the last term
is of the same order as the other terms only if the contracted indices 
belong to the same Kronecker delta and has to be dropped in all other cases.

\subsection{Solution of the leading order equations for the spin field
operator}

In this subsection the number of particles, $r$, is an odd number and
we will use the notation $\nu=(r-1)/2$. For the leading order form
factor we take the following ansatz
\begin{equation} 
F^x_{a_1\dots a_r}(\theta_1,\dots,\theta_r)={\cal N}_r\sum_\sigma\,
\delta^x_{a_{\sigma(1)}}\,R^\sigma_{23}\cdots
R^\sigma_{r-1\,r}\,Q_r(\theta_{\sigma(1)},\dots,\theta_{\sigma(r)})\,,
\label{ansatz1}
\end{equation} 
where
\begin{equation} 
{\cal N}_r=\frac{1}{n^\nu}\,\frac{1}{2^\nu\nu!}\,
\left(\frac{i}{2}\right)^{\nu(\nu-1)}
\end{equation} 
and $\sigma$ runs over the $r!$ permutations of the particles. Finally
we used the shorthand notation ($\epsilon$ is the sign function)
\begin{equation} 
R^\sigma_{ij}=\delta_{a_{\sigma(i)}a_{\sigma(j)}}\,
R\left[\epsilon\left(\sigma(j)-\sigma(i)\right)(\theta_{\sigma(i)}-
\theta_{\sigma(j)})\right]\,,
\end{equation} 
which for physical (real, ordered) rapidities reduces to
\begin{equation} 
\delta_{a_{\sigma(i)}a_{\sigma(j)}}\,
R\left[\vert\theta_{\sigma(i)}-\theta_{\sigma(j)}\vert\right]\,.
\end{equation} 
We require that the scalar function $Q_r(\theta_1,\dots,\theta_r)$
is symmetric under the exchanges $2\leftrightarrow3$, 
$4\leftrightarrow5$,$\dots$,$r-1\leftrightarrow r$ and is totally
symmetric under permutation of these pairs of variables. Further we
require that $Q_r$ is $2\pi i$ periodic in all variables and is even
under simultaneous sign change of all variables. Then it is almost
obvious that the ansatz (\ref{ansatz1}) satisfies the homogeneous
equations (\ref{NAX1}--\ref{NAX5})\footnote{Note the relation
$R(\theta)=R(-\theta)[1+a_1(\theta)]$.}. 
It is also possible to write
\begin{eqnarray}
&&F^x_{a_1\dots a_r}(\theta_1,\dots,\theta_r)\\&=&\frac{1}{n^\nu}\,
\left(\frac{i}{2}\right)^{\nu(\nu-1)}\,\delta^x_{a_1}\,\delta_{a_2a_3}\,
R(\theta_2-\theta_3)\cdots
\delta_{a_{r-1}a_r}\,R(\theta_{r-1}-\theta_r)\,Q_r(\theta_1,\dots,
\theta_r)+\cdots,\nonumber\\
\nonumber
\end{eqnarray}
where the final dots stand for all similar terms, corresponding to such
permutations of the variables not leaving the first term invariant.
Finally (\ref{NAX4}) will also
be satisfied by (\ref{ansatz1}) if the set of scalar functions $Q_r$
obeys the following three relations for $r\geq3$
\begin{equation} 
\begin{split} 
Q_{r+2}&(\theta_1,\beta+i\pi,\beta,\theta_2,\dots,\theta_r)\\
&=\left\{\prod_{j=1}^r\sinh(\beta-\theta_j)\right\}\,
\left\{\sum_{k=1}^r\frac{1}{\sinh(\beta-\theta_k)}\right\}\,
Q_r(\theta_1,\dots,\theta_r)\,,
\end{split}
\label{rec1}
\end{equation} 
\begin{equation} 
Q_{r+2}(\beta+i\pi,\beta,\theta_1,\dots,\theta_r)
=\left\{\prod_{j=2}^r\sinh(\beta-\theta_j)\right\}\,
Q_r(\theta_1,\dots,\theta_r)\,,
\label{rec2}
\end{equation} 
\begin{equation}
\begin{split} 
Q_{r+2}(\theta_1,&\beta+i\pi,\theta_2,\beta,\theta_3,\dots,\theta_r)\\
&=\sinh(\theta_2-\theta_3)\,\left\{\prod_{j\not=2,3}^r
\sinh(\beta-\theta_j)\right\}\,
Q_r(\theta_1,\dots,\theta_r)\,.
\end{split} 
\label{rec3}
\end{equation} 

Using the results of Subsect.~\ref{3FFn} we see that
\begin{equation} 
Q_3(\theta_1,\theta_2,\theta_3)=1\,.
\label{q3}
\end{equation} 
For $r=5$ we have
\begin{eqnarray}
%\begin{split} 
&&Q_5(\theta_1,\theta_2,\theta_3,\theta_4,\theta_5)=
-\frac{1}{2}\Big[1+
\sum_{k<l}\cosh(\theta_k-\theta_l)\nonumber\\
&&+\cosh(\theta_2+\theta_3-\theta_4-\theta_5)
+\cosh(\theta_1+\theta_3-\theta_4-\theta_5)\\
&&+\cosh(\theta_2+\theta_3-\theta_1-\theta_5)
+\cosh(\theta_2+\theta_3-\theta_4-\theta_1)
+\cosh(\theta_2+\theta_1-\theta_4-\theta_5)\Big]\,.\nonumber\\
\nonumber
%\end{split}
\label{q5}
\end{eqnarray} 
It is easy to check that this satisfies (\ref{rec1}--\ref{rec3}) for 
$r=3$. Because of the $2\pi i$ periodicity the scalar function $Q_r$ is
really a function of the exponential variables $x_k={\rm e}^{\theta_k}$,
$k=1,2,\dots,r$.
Using the fact that the form factors are regular functions that are
also regular at infinity and also taking into account the presence of
the denominator in (\ref{reducedF}) we can show that 
$Q_r$, as function of one of the variables, say $x_2$, is a finite Laurent
polynomial consisting of the terms $x_2^{\nu-1}, x_2^{\nu-2},\dots,
x_2^{1-\nu}$. Applying this to the $r+2$ case, we see that
$x_2^\nu Q_{r+2}(\theta_1,\dots,\theta_{r+2})$ is a polynomial of degree
$r-1$ in $x_2$ hence it is determined by its values at $r$ different
points. If $Q_r$ is given, these data are provided by
(\ref{rec1}) and (\ref{rec3}) and we can use them as recursion relations
to determine $Q_{r+2}$. The solution is given by the explicit formula
\begin{equation} 
\begin{split}
Q_{r+2}(&\theta_1,\theta_2,\theta_3,\dots,\theta_{r+2})\\
&=\sum_{k=3}^{r+2}\,Q_{r+2}(\theta_1,\theta_k+i\pi,\theta_3,\dots,\theta_{r+2})
\,\left(-\frac{x_k}{x_2}\right)^{\frac{r-1}{2}}\,\prod_{l\not=1,2,k}\,
\frac{x_l+x_2}{x_l-x_k}\,.
\end{split}
\label{REC}
\end{equation} 
We have applied (\ref{REC}) to determine $Q_7$. We have checked that 
the function $Q_7$ constructed this way is also a polynomial in all the
other variables and satisfies all symmetry requirements together with
(\ref{rec2}) (which was not used in the construction (\ref{REC})).
It would be interesting to show analogous results for general $r$.

\subsection{Noether current and symmetric tensor form factors} 

The leading order form factors of the Noether current and the
symmetric, traceless isotensor operator are given by the ansatz
\begin{equation} 
F^{xy}_{a_1\dots a_r}(\theta_1,\dots,\theta_r)={\cal M}_r\sum_\sigma\,
T^\sigma\,R^\sigma_{34}\cdots
R^\sigma_{r-1\,r}\,Q_r(\theta_{\sigma(1)},\dots,\theta_{\sigma(r)})\,,
\label{ansatz2}
\end{equation} 
where $r$ is even, $\mu=(r-2)/2$,
\begin{equation} 
{\cal M}_r=\left(\frac{i}{n}\right)^\mu\,\frac{1}{2^{\mu+1}\mu!}\,
\left(\frac{i}{2}\right)^{\mu^2}
\end{equation} 
and 
\begin{equation} 
T^\sigma=T^{xy}_{a_{\sigma(1)}a_{\sigma(2)}}
(\theta_{\sigma(1)}-\theta_{\sigma(2)})\,,
\end{equation} 
where
\begin{equation} 
T^{xy}_{a_1a_2}(\theta)=\left\{
\begin{split}&t^{xy}_{a_1a_2}\sinh\theta\quad\ \,{\rm (current),}\\
&s^{xy}_{a_1a_2}\qquad\quad\quad {\rm (tensor).}\\
\end{split}\right.
\end{equation} 

For even $r$ the scalar function $Q_r(\theta_1,\dots,\theta_r)$
is symmetric under the exchanges $1\leftrightarrow2$, 
$3\leftrightarrow4$,$\dots$,$r-1\leftrightarrow r$ and is totally
symmetric under permutation of the last $\mu$ pairs of variables. 
Further $Q_r$ is $2\pi i$ antiperiodic in all variables and is invariant
under simultaneous sign change of all variables. 
The ansatz (\ref{ansatz2}) with $Q_r$ satisfying the above symmetry 
requirements satisfies the homogeneous equations (\ref{NAX1}--\ref{NAX5}). 
We introduce the functions $P^{(o)}_r$ for $o=c,t$
\begin{equation} 
Q_r(\theta_1,\dots,\theta_r)=\left\{
\begin{split}
P^{(c)}_r(\theta_1,\dots,\theta_r)\qquad\quad&{\rm (current),}\\
2\cosh\left(\frac{\theta_1-\theta_2}{2}\right)\,
P^{(t)}_r(\theta_1,\dots,\theta_r)\quad&{\rm (tensor).}\\
\end{split}\right.
\end{equation} 
(\ref{ansatz2}) also satisfies the residue equation (\ref{NAX4}) if
for $r\geq4$
\begin{equation}
\begin{split} 
P^{(o)}_{r+2}(\theta_1,\theta_2,&\beta+i\pi,\theta_3,\beta,\theta_4,
\dots,\theta_r)\\
&=-i\sinh(\theta_3-\theta_4)\,\left\{\prod_{j\not=3,4}
\sinh(\beta-\theta_j)\right\}\,
P^{(o)}_r(\theta_1,\dots,\theta_r)\,,
\end{split} 
\label{REC1}
\end{equation}
and for $r\geq2$ 
\begin{equation} 
\begin{split} 
P^{(o)}_{r+2}&(\theta_1,\theta_2,\beta+i\pi,\beta,\theta_3,\theta_4,
\dots,\theta_r)\\
&=-i\left\{\prod_{j=1}^r\sinh(\beta-\theta_j)\right\}\,
\left\{\sum_{k=1}^r\frac{1}{\sinh(\beta-\theta_k)}\right\}\,
P^{(o)}_r(\theta_1,\dots,\theta_r)\,,
\end{split}
\label{REC2}
\end{equation} 
\begin{equation}
\begin{split} 
P^{(c)}_{r+2}(&\beta+i\pi,\theta_1,\beta,\theta_2,
\dots,\theta_r)\\
&=-i\sinh(\theta_1-\theta_2)\,\left\{\prod_{j=3}^r
\sinh(\beta-\theta_j)\right\}\,
P^{(c)}_r(\theta_1,\dots,\theta_r)\,,
\end{split} 
\label{REC3}
\end{equation} 
\begin{equation}
\begin{split} 
&P^{(t)}_{r+2}(\beta+i\pi,\theta_1,\beta,\theta_2,
\dots,\theta_r)\\
&=-2\cosh\left(\frac{\theta_1-\theta_2}{2}\right)\,
\cosh\left(\frac{\beta-\theta_1}{2}\right)\,
\left\{\prod_{j=3}^r
\sinh(\beta-\theta_j)\right\}\,
P^{(t)}_r(\theta_1,\dots,\theta_r)\,.
\end{split} 
\label{REC4}
\end{equation} 
The $P^{(o)}_2$ functions are given by
\begin{equation} 
P^{(c)}_2(\theta_1,\theta_2)=
\frac{-1}{2\cosh\left(\frac{\theta_1-\theta_2}{2}\right)},\qquad\qquad
P^{(t)}_2(\theta_1,\theta_2)=
\frac{1}{2}
\end{equation} 
and for $r=4$ we have 
\begin{eqnarray} 
P^{(c)}_4(\theta_1,\theta_2,\theta_3,\theta_4)&=&
\cosh\left(\frac{\theta_1+\theta_2-\theta_3-\theta_4}{2}\right)\,,\\
P^{(t)}_4(\theta_1,\theta_2,\theta_3,\theta_4)&=&
-\cosh\left(\frac{\theta_1-\theta_3}{2}\right)
\cosh\left(\frac{\theta_1-\theta_4}{2}\right)\\
&&-\cosh\left(\frac{\theta_2-\theta_3}{2}\right)
\cosh\left(\frac{\theta_2-\theta_4}{2}\right)\,.\nonumber\\
\nonumber
\end{eqnarray} 

Similarly to the case of field operators discussed in the previous
subsection, the expression $x_3^{\frac{r-1}{2}}\,P^{(o)}_{r+2}$,
treated as a function of the variable $x_3$, is a polynomial of degree
$r-1$ and is determined by its values at $r$ different points. The
recursion relations (\ref{REC1}--\ref{REC4}) can be used to determine 
this expression at $r$ different points and the Laurent polynomials
$P^{(o)}_{r+2}$ can be calculated from a formula similar to (\ref{REC}).

%\vfill
%\eject

\section{\lan of the functional integral}

In this section we check that the bootstrap solutions for the
multi--particle form factors found in leading order $1/n$ expansion
in Subsections~5.5,~5.6 do indeed correspond to those obtained by
quantum field theoretic calculations 
\footnote{Checks of the S--matrix itself
to leading orders in $1/n$ were performed much earlier \cite{BKKW}.}.
 
The $1/n$ expansion of the functional integral of the \On non--linear
$\sigma$--model has been described in numerous papers. Starting from
bare fields $q^a$ one imposes the constraint $q^2=n$ by introducing
a Lagrange multiplier field $\lambda$. Here we just recall the 
resulting Feynman rules for computation of the correlation functions of
the elementary field:
\ba
q\,\,{\rm propagator}:\,&=&\,\delta^{ab}iD(p,m_0)\,,\,\,\,\,\,\,
D(p,m_0)=\frac{1}{p^2-m_0^2+i\epsilon}\,,
\\
\lambda\,\,{\rm propagator}:\,&=&\,2J(p,m_0)^{-1}\,,
\\
q\,-\lambda\,\,{\rm vertex}:\,&=&\,\frac{1}{\sqrt{n}}\delta^{ab}
\ea
\noindent with momentum conservation at each vertex

\noindent -- for each external line a factor $Z^{-1/2}$

\noindent -- for each closed $q$--loop there is a factor $n$

\noindent -- only $q$--loops with more than 2 vertices should be drawn

\noindent -- integration $\int\frac{\rmd^2k}{(2\pi)^2}$ over all internal 
momenta $k$ for which a cutoff $\Lambda$ is imposed 
(e.g. Pauli--Villars for the $q$--propagator).

Renormalization of the bare parameters order by order in $1/n$ is given by
\ba
m_0^2&=&M^2\left(1-\sum_{s=1}^{\infty}\frac{\alpha_s}{n^s}\right)
\,\,\,\,\,; \alpha_s=\alpha_s(\Lambda/M)\,,
\\
Z&=&1+\sum_{s=1}^{\infty}\frac{Z_s}{n^s}\,\,\,\,\,; Z_s=Z_s(\Lambda/M)\,.
\ea

The $\lambda$ inverse propagator function $J(q,m)$ is a special case
of the 1--loop integrals:
\be
J_r(q_1,\dots,q_r,m)=\int\frac{\rmd^2k}{(2\pi)^2}
\prod_{j=1}^r D(k+l_j,m)\,,
\end{equation}
where
\ba
q_j&=&l_j-l_{j-1}\,,\,\,\,\,l_{-1}=l_r\,,
\\
\sum_{j=1}^r q_j&=&0\,,
\ea
which can be computed using the cutting rules. In particular we have
\be
J(q,m)\equiv J_2(q,-q,m)=\frac{i}{4m^2 R(\theta)}
\,\,\,\,\,{\rm for}\,\,q^2=4m^2\cosh^2\left(\frac{\theta}{2}\right)\,.
\end{equation}
We will also need the case $r=3$:
\ba
J_3(q_1,q_2,q_3,m)&=&-\frac{(q_1q_2)q_3^2 J(q_3,m)}{D_3(q_1,q_2,q_3,m)}
+2\,\,{\rm perms}\,,
\label{j3id}
\\
D_3(q_1,q_2,q_3,m)&=&q_1^2q_2^2q_3^2+m^2\lambda(q_1^2,q_2^2,q_3^2)
-i\epsilon q_1q_2\,,
\\
\lambda(x_1,x_2,x_3)&=&x_1^2+x_2^2+x_3^2-2x_1x_2-2x_2x_3-2x_3x_1\,.
\ea

\subsection{3--particle spin form factor}

The 3--particle form factor in leading order is, using the Feynman rules
above, simply obtained from a sum of tree graphs and amputating three
of the external lines thereby obtaining:
\be
f^a_{b_1b_2b_3}(\theta_1,\theta_2,\theta_3)=\frac{2}{n}\frac{i}{q^2-M^2}
\left[\delta^a_{b_1}\delta_{b_2b_3}J(p_2+p_3)^{-1}
+2\,\,{\rm perms}\right]
+\rmO(1/n^2)\,,
\end{equation}
where $q=p_1+p_2+p_3$. Here and in the rest of this section we  
omit the argument $M$ in the functions $J_r,D$ i.e. $J(s)=J(s,M)$. 
For three incoming (on--shell $p_i^2=M^2$) particles  one has
\be
q^2-M^2=8M^2 C_3(\theta_1,\theta_2,\theta_3)\,,
\end{equation}
thus producing Eq.~(\ref{ansatz1}) for the case $r=3$ with (\ref{q3}).

\subsection{4--particle current and isotensor form factors}

Consider the current
\be
J_\mu^{ab}=q^a\partial_\mu q^b-q^b\partial_\mu q^a
\end{equation}
whose 2--particle form factor is in leading order just given by the 
contact diagram ($q=p_1+p_2$): 
\be
-i\epsilon_{\mu\alpha}q^\alpha f^{ab}_{b_1b_2}(\theta_1,\theta_2)
=it^{ab}_{b_1b_2}(p_1-p_2)_\mu+\rmO(1/n)\,,
\end{equation}
yielding 
$f^{ab}_{b_1b_2}(\theta_1,\theta_2)
=-t^{ab}_{b_1b_2}\tanh\frac{(\theta_1-\theta_2)}{2}$ 
as required.

The 4--particle current form factor in leading order is a 
sum of tree diagrams:
\ba
&&-i\epsilon_{\mu\nu}q^\nu f^{ab}_{b_1b_2b_3b_4}
(\theta_1,\theta_2,\theta_3,\theta_4)
=-\frac{2}{n}\sum_{1\le i<j\le4}
t^{ab}_{b_ib_j}\delta_{b_kb_l}V_\mu(q,p_i,p_j)
J\left(p_k+p_l\right)^{-1}
\nonumber\\
&&+\rmO(1/n^2)\,,
\ea
where $k<l$ and $\{i,j\}\cup\{k,l\}=\{1,2,3,4\}\,,\,\,$
$q=\sum_{j=1}^4 p_j$, and
\be
V_\mu(q,p_i,p_j)\equiv(2p_i-q)_\mu D(q-p_i)-(2p_j-q)_\mu D(q-p_j)\,.
\end{equation}
For on--shell momenta $p_i$ it is clear
that $q^\mu V_\mu(q,p_i,p_j)=0$ as required for current conservation,
and one can check that
\be
V_\mu(q,p_i,p_j)=\frac{1}{16M^2}\epsilon_{\mu\nu}q^\nu\, 
\frac{\sinh(\theta_i-\theta_j)\cosh\frac12
\left(\theta_i+\theta_j-\theta_k-\theta_l\right)}
{C_4(\theta_1,\theta_2,\theta_3,\theta_4)}\,,
\end{equation}
thus deriving the 4--particle bootstrap solution in subsect.~5.6.

Similarly for the 4--particle isotensor form factor:
\ba
&&\tilde{f}^{ab}_{b_1b_2b_3b_4}(\theta_1,\theta_2,\theta_3,\theta_4)
=\frac{2i}{n}\sum_{1\le i<j\le4}
s^{ab}_{b_ib_j}\delta_{b_kb_l}
\left[D(q-p_i)+D(q-p_j)\right]
\nonumber\\
&&\times J\left(p_k+p_l,m\right)^{-1}+\rmO(1/n^2)\,.
\ea
For on--shell momenta $p_i$ one has:
\ba
&&D(q-p_i)+D(q-p_j)=\frac{1}{8M^2}
\frac{\cosh\left(\frac{\theta_i-\theta_j}{2}\right)}
{C_4(\theta_1,\theta_2,\theta_3,\theta_4)}\times
\nonumber\\
&&\left[
 \cosh\left(\frac{\theta_i-\theta_k}{2}\right)
 \cosh\left(\frac{\theta_i-\theta_l}{2}\right)
+\cosh\left(\frac{\theta_j-\theta_k}{2}\right)
 \cosh\left(\frac{\theta_j-\theta_l}{2}\right)\right],
\ea
again consistent with the corresponding result in Subsect.~5.6.

\subsection{5--particle spin field form factor}

The leading contribution to the 5--particle (in state) spin form factor
is $\rmO(1/n^2)$. It is a little more complicated since there
are two types of diagrams contributing: tree diagrams with two
$\lambda$--propagators and others involving a closed $q$--triangle 
connected to the external lines by three 
$\lambda$--propagators. Using the $1/n$ rules outlined above 
one gets ($q=\sum_{j=1}^5 p_j$):
\ba
&&f^a_{b_1b_2b_3b_4b_5}(\theta_1,\theta_2,\theta_3,\theta_4,\theta_5)
=-\frac{4}{n^2}\frac{1}{(q^2-M^2)}
\Bigl[
\nonumber\\
&&\delta^c_{b_1}\delta_{b_2b_3}\delta_{b_4b_5}
f^{(5)}(\theta_1,\theta_2,\theta_3,\theta_4,\theta_5)
+14\,\,{\rm perms}\Bigr]+\rmO(1/n^3)\,,
\ea
where
\ba
&&f^{(5)}(\theta_1,\theta_2,\theta_3,\theta_4,\theta_5)
=\frac{\left\{D(p_1+p_4+p_5)+D(p_1+p_2+p_3)\right\}}{J(p_2+p_3)J(p_4+p_5)}
\nonumber\\
&&+\frac{\left\{D(p_2+p_3+p_4)+D(p_2+p_3+p_5)\right\}}{J(q-p_1)J(p_2+p_3)}
+\frac{\left\{D(p_2+p_4+p_5)+D(p_3+p_4+p_5)\right\}}{J(q-p_1)J(p_4+p_5)}
\nonumber\\
&&-2\frac{J_3(p_1-q,p_2+p_3,p_4+p_5)}{J(q-p_1)J(p_2+p_3)J(p_4+p_5)}\,.
\ea
Using (\ref{j3id}) this can be rewritten
\be
f^{(5)}_1(\underline{\theta})=\frac{U_1}{J(q_2)J(q_3)}+\frac{U_2}{J(q_1)J(q_3)}
+\frac{U_3}{J(q_1)J(q_2)}\,,
\end{equation}
with
\ba
U_1&=&D(p_1+q_2)+D(p_1+q_3)+\frac{2(q_2q_3)q_1^2}{D_3(q_1,q_2,q_3)}\,,
\\
U_2&=&D(p_2+q_3)+D(p_3+q_3)+\frac{2(q_1q_3)q_2^2}{D_3(q_1,q_2,q_3)}\,,
\\
U_3&=&D(p_4+q_2)+D(p_5+q_2)+\frac{2(q_1q_2)q_3^2}{D_3(q_1,q_2,q_3)}\,,
\ea
where
\be
q_1=p_1-q\,,\,\,\,\,\,q_2=p_2+p_3\,,\,\,\,\,\,q_3=p_4+p_5\,.
\end{equation}
We note
\be
D_3(q_1,q_2,q_3)=256M^6C_4(\theta_2,\theta_3,\theta_4,\theta_5)
\cosh\left(\frac{\theta_2-\theta_3}{2}\right)
\cosh\left(\frac{\theta_4-\theta_5}{2}\right)\,,
\end{equation}
and then after some algebra it can be shown that 
\be
U_2=0=U_3\,,
\end{equation}
and \footnote{The vanishing of $U_1$ as $q$ goes on-shell is consistent
with the absence of particle production.}
\be
U_1=-\frac{(q^2-M^2)Q_5(\theta_1,\theta_2,\theta_3,\theta_4,\theta_5)}
{256 M^4 C_5(\theta_1,\theta_2,\theta_3,\theta_4,\theta_5)}\,,
\end{equation}
with $Q_5$ defined in Eq.~(\ref{q5}), thus reproducing the result in
Subsect.~5.5. 

\subsection{6--particle current and isotensor form factors}

Inspecting the diagrams contributing to the  
6--particle current form factor in the leading order $1/n$ expansion
we note that we can write these in terms of
form factors of the spin field (here $q=\sum_{j=1}^6p_j$): 
\ba
&&-i\epsilon_{\mu\nu}q^\nu f^{ab}_{b_1b_2b_3b_4b_5b_6}
(\theta_1,\theta_2,\theta_3,\theta_4,\theta_5,\theta_6)
\nonumber\\
&=&i\left[t^{ab}_{b_1c}(2p_1-q)_\mu 
f^c_{b_2b_3b_4b_5b_6}(\theta_2,\theta_3,\theta_4,\theta_5,\theta_6)
+5\,\,{\rm similar\,\,terms}\right]
\nonumber\\
&+&i\Bigl[t^{ab}_{cd}(p_1+p_2+p_3-p_4-p_5-p_6)_\mu 
f^c_{b_1b_2b_3}(\theta_1,\theta_2,\theta_3)
f^d_{b_4b_5b_6}(\theta_4,\theta_5,\theta_6)
\nonumber\\
&+&9\,\,{\rm similar\,\,terms}\Bigr]+\rmO(1/n^3)\,.
\ea
Using the expressions previously obtained we get
\ba
&&-i\epsilon_{\mu\nu}q^\nu f^{ab}_{b_1b_2b_3b_4b_5b_6}
(\theta_1,\theta_2,\theta_3,\theta_4,\theta_5,\theta_6)
=-\frac{i}{4n^2}\Bigl\{t^{ab}_{b_1b_2}\delta_{b_3b_4}\delta_{b_5b_6}
R(\theta_{34})R(\theta_{56})
\nonumber\\
&&\times\Bigl[
 \frac{(2p_1-q)_\mu Q_5(\theta_2,\theta_3,\theta_4,\theta_5,\theta_6)}
{C_5(\theta_2,\theta_3,\theta_4,\theta_5,\theta_6)}
-\frac{(2p_2-q)_\mu Q_5(\theta_1,\theta_3,\theta_4,\theta_5,\theta_6)}
{C_5(\theta_1,\theta_3,\theta_4,\theta_5,\theta_6)}
\nonumber\\
&-&4\left(\frac{(p_1+p_3+p_4-p_2-p_5-p_6)_\mu}
{C_3(\theta_1,\theta_3,\theta_4)C_3(\theta_2,\theta_5,\theta_6)}
-\frac{(p_2+p_3+p_4-p_1-p_5-p_6)_\mu}
{C_3(\theta_2,\theta_3,\theta_4)C_3(\theta_1,\theta_5,\theta_6)}
\right)\Bigr]
\nonumber\\
&+&44\,\,{\rm similar\,\,terms}\Bigr\}+\rmO(1/n^3)\,.
\ea
One can check that contracting the rhs with $q_\mu$ is zero 
and then obtain the representation
for the 6--particle current form factor with $P_6^{(c)}$
given in Appendix~C.

Similarly for the isotensor:
\ba
&&\tilde{f}^{ab}_{b_1b_2b_3b_4b_5b_6}
(\theta_1,\theta_2,\theta_3,\theta_4,\theta_5,\theta_6)
\nonumber\\
&=&\left[s^{ab}_{b_1c} 
f^c_{b_2b_3b_4b_5b_6}(\theta_2,\theta_3,\theta_4,\theta_5,\theta_6)
+5\,\,{\rm similar\,\,terms}\right]
\nonumber\\
&+&\Bigl[s^{ab}_{cd} 
f^c_{b_1b_2b_3}(\theta_1,\theta_2,\theta_3)
f^d_{b_4b_5b_6}(\theta_4,\theta_5,\theta_6)
+9\,\,{\rm similar\,\,terms}\Bigr]+\rmO(1/n^3)\,.
\\
&&
\nonumber\\
&=&-\frac{1}{4n^2}\Bigl\{s^{ab}_{b_1b_2}\delta_{b_3b_4}\delta_{b_5b_6}
R(\theta_{34})R(\theta_{56})
\nonumber\\
&&\times\Bigl[
 \frac{Q_5(\theta_2,\theta_3,\theta_4,\theta_5,\theta_6)}
{C_5(\theta_2,\theta_3,\theta_4,\theta_5,\theta_6)}
+\frac{Q_5(\theta_1,\theta_3,\theta_4,\theta_5,\theta_6)}
{C_5(\theta_1,\theta_3,\theta_4,\theta_5,\theta_6)}
\nonumber\\
&-&4\left(
 \frac{1}{C_3(\theta_1,\theta_3,\theta_4)C_3(\theta_2,\theta_5,\theta_6)}
+\frac{1}{C_3(\theta_2,\theta_3,\theta_4)C_3(\theta_1,\theta_5,\theta_6)}
\right)\Bigr]
\nonumber\\
&+&44\,\,{\rm similar\,\,terms}\Bigr\}+\rmO(1/n^3)\,.
\ea
The expression for $P_6^{(t)}$ thus obtained agrees with the
corresponding bootstrap solution as expected.

%\vfill
%\eject

\section{Large $n$ clustering tests}

\subsection{$3\,\rightarrow\, 2+1$ clustering in the large $n$ expansion}

We have already computed the 2--particle and 3--particle form factors
in the first two orders of the large $n$ expansion
so we are able to check the clustering formula (\ref{321clu})
in this limit.

For the 2--particle form factors we have
\be
\psi_I(\theta)=F_I(\theta)+\frac{1}{n}\,G_I(\theta)+\dots \qquad
I=1,2,
\end{equation}
where
\be
F_1(\theta)=\tanh\frac{\theta}{2},\qquad\qquad\quad F_2(\theta)=i
\end{equation}
and
\be
G_1(\theta)=2\pi\tanh\frac{\theta}{2}\left[a(\theta)+b(\theta)\right],
\qquad\qquad\quad G_2(\theta)=2\pi i\left[a(\theta)-b(\theta)\right]\,.
\end{equation}

In the case of the 3--particle form factors we first write
\be
\begin{split}
f^x_{abc}(\Delta+\beta,\beta,\gamma)&\cong
\frac{8\rme^{-\Delta}}{\rme^\theta+1}\,\Big\{
\delta^x_a\delta_{bc}\,X(\Delta,\Delta+\theta)\\
&+\delta^x_b\delta_{ac}\,X(\Delta,2\pi i-\theta)+
\delta^x_c\delta_{ab}\,X(\theta,\Delta+\theta)\Big\}\,,
\end{split}
\label{3X}
\end{equation}
where $\theta=\beta-\gamma$.
From here it is easy to calculate that the coefficient of the 
leading ($1/n$) term on the right hand side of (\ref{3X}) is
\be
-\frac{2\pi}{\Delta}\,\left\{F_1(\theta)t^{xa}_{bc}-i
F_2(\theta)s^{xa}_{bc}\right\}+\dots
\end{equation}
Comparing it to (\ref{321clu0}) we see that
\be
H_2=1+\chi\omega+\rmO(\chi^2)\,,
\label{H20}
\end{equation}
where $\omega$ is a real number.
Using this result and (\ref{321clu0}) the predicted form of the
NLO ($1/n^2$) coefficient on the right hand side of (\ref{3X}) becomes
\be
-\frac{2\pi}{\Delta}\left\{\left(2F_1+G_1\right)t^{xa}_{bc}
-is^{xa}_{bc}\left(
2F_2+G_2-2F_2\,\ln\frac{\Delta}{2\pi}+\omega F_2\right)+
2iF_2\delta^x_a\delta_{bc}\right\}\,.
\label{second}
\end{equation}
(Here the argument of all the functions $F_I$, $G_I$ is $\theta$.)

Now we use the large $\Delta$ asymptotic expansions
\be
\begin{split}
g(\Delta,\Delta+\theta)&\cong
\frac{\pi}{2\Delta}\rme^\Delta\left(1+\rme^\theta\right)\\
g(\Delta,2\pi i-\theta)&\cong-\frac{\pi}{2\Delta}\rme^\Delta
\Bigg\{\rme^\theta\left[2+2\pi a(\theta)\right]\\
&+\left(\rme^\theta+1\right)\left[\Psi\left(\frac{1}{2}\right)-
\ln\left(\frac{\Delta}{2\pi}\right)\right]-2\pi b(\theta)\Bigg\}\\
g(\theta,\Delta+\theta)&\cong-\frac{\pi}{2\Delta}\rme^\Delta
\Bigg\{\left[2+2\pi a(\theta)\right]\\
&+\left(\rme^\theta+1\right)\left[\Psi\left(\frac{1}{2}\right)-
\ln\left(\frac{\Delta}{2\pi}\right)\right]-2\pi b(\theta)
\rme^\theta\Bigg\}
\end{split}
\end{equation}
and calculate the large $\Delta$ asymptotics of our NLO large $n$
3--particle form factor explicitly. We find that it is indeed exactly
of the form (\ref{second}), if we choose 
\be
\omega=2\Psi\left(\frac{1}{2}\right)\,.
\end{equation}

\subsection{Odd--odd clustering}

Specifying the odd--odd clustering formula for the current form factors
at the leading order in the large $n$ expansion gives
\begin{equation}
\begin{split}
&F^{cd}_{a_1\dots a_kb_1\dots b_l}(\alpha_1+\Delta,\dots,
\alpha_k
+\Delta,\beta_1,\dots,\beta_l)
\cong -\left(\frac{\rme^{\Delta/2}}{2}\right)^{kl}\,
t^{cd}_{ab}\\
&\cdot\exp\left\{\frac{l}{2}\sum_{i=1}^k\alpha_i
-\frac{k}{2}\sum_{j=1}^l\beta_j\right\}\,
F^a_{a_1\dots a_k}(\alpha_1,\dots,\alpha_k)\,
F^b_{b_1\dots b_l}(\beta_1,\dots,\beta_k)\,.
\end{split}
\label{Noo}
\end{equation}
Note the appearance of prefactors which arise because we are considering 
reduced form factors here.

Analogously for the leading large $n$ form factors of the symmetric
tensor we get
\begin{equation}
\begin{split}
&\tilde F^{cd}_{a_1\dots a_kb_1\dots b_l}(\alpha_1+\Delta,\dots,
\alpha_k
+\Delta,\beta_1,\dots,\beta_l)
\cong \left(\frac{\rme^{\Delta/2}}{2}\right)^{kl}\,
s^{cd}_{ab}\\
&\cdot\exp\left\{\frac{l}{2}\sum_{i=1}^k\alpha_i
-\frac{k}{2}\sum_{j=1}^l\beta_j\right\}\,
F^a_{a_1\dots a_k}(\alpha_1,\dots,\alpha_k)\,
F^b_{b_1\dots b_l}(\beta_1,\dots,\beta_k)\,.
\end{split}
\label{Noo2}
\end{equation}

Using the explicit solution of the current and symmetric tensor as
well as the spin field form factors we can prove that the $k=1$
special case of the clustering relations implies 
\be
%\begin{split}
K_l(\beta_1,\beta_2,\dots,\beta_l)=-
\,\left(-\frac{1}{2}\right)^{\frac{l-1}{2}}
\exp\left\{\frac{1}{2}\beta_1
-\frac{1}{2}\sum_{j=2}^l\beta_j\right\}\,
Q_l(\beta_1,\beta_2,\dots,\beta_l)
\label{Kl}
\end{equation}
and
\be
%\begin{split}
L_l(\beta_1,\beta_2,\dots,\beta_l)=-
\,\left(-\frac{1}{2}\right)^{\frac{l+1}{2}}
\exp\left\{-\frac{1}{2}\sum_{j=2}^l\beta_j\right\}\,
Q_l(\beta_1,\beta_2,\dots,\beta_l)\,.
\label{Ll}
\end{equation}
Here the functions $K_l$ and $L_l$ are defined by the asymptotic relations
\be
P^{(c)}_{l+1}(\Delta,\beta_1,\dots,\beta_l)\cong\exp\left\{
\frac{\Delta(l-2)}{2} \right\}\,
K_l(\beta_1,\dots,\beta_l)
\end{equation}
and 
\be
P^{(t)}_{l+1}(\Delta,\beta_1,\dots,\beta_l)\cong\exp\left\{
\frac{\Delta(l-1)}{2} \right\}\,
L_l(\beta_1,\dots,\beta_l)\,.
\end{equation}
We have checked the clustering relations (\ref{Kl}) and (\ref{Ll})
for $l=3,5$ explicitly.

\subsection{Odd--even clustering}

Using the general odd--even clustering formula we can calculate the
reduced form factor clustering in the large $n$ expansion at leading order: 
\begin{equation}
\begin{split}
&F^x_{a_1\dots a_kb_1\dots b_l}(\alpha_1+\Delta,\dots,
\alpha_k
+\Delta,\beta_1,\dots,\beta_l)\\
&\cong \frac{2\pi}{n\Delta}\,\left(\frac{\rme^{\Delta/2}}{2}\right)^{kl}\,
\exp\left\{\frac{l}{2}\sum_{i=1}^k\alpha_i
-\frac{k}{2}\sum_{j=1}^l\beta_j\right\}\,
F^a_{a_1\dots a_k}(\alpha_1,\dots,\alpha_k)\\
&\qquad\qquad\cdot \left\{F^{xu}_{b_1\dots b_l}(\beta_1,\dots,\beta_l)-
\tilde F^{xu}_{b_1\dots b_l}(\beta_1,\dots,\beta_l)\right\}\,.
\end{split}
\label{Noe}
\end{equation}
(Here $k$ is odd and $l$ is even.)

We define the function $M_l$ by the asymptotic formula
\be
Q_{l+1}(\beta_1,\Delta,\beta_2,\dots,\beta_l)\cong\exp\left\{
\frac{\Delta(l-2)}{2} \right\}\,
M_l(\beta_1,\beta_2,\dots,\beta_l)\,.
\end{equation}
In the special case $k=1$ (\ref{Noe}) leads to the following recursion 
relation:
\be
\begin{split}
&\rme^{-\beta_2}\,M_l(\beta_1,\beta_2,\dots,\beta_l)=
\left(\frac{1}{2}\right)^{\frac{l-2}{2}}
\exp\left\{-\frac{1}{2}\sum_{j=1}^l\beta_j\right\}\\
&\cdot\left\{2\cosh\left(\frac{\beta_1-\beta_2}{2}\right)\,
P^{(t)}_l(\beta_1,\dots,\beta_l)-\sinh(\beta_1-\beta_2)\,
P^{(c)}_l(\beta_1,\dots,\beta_l)\right\}\,.
\end{split}
\label{Ml}
\end{equation}
We have checked this relation for $l=2,4,6$.

Similarly the $l=2$ special case corresponds to the 
recursion relation
\be
N_k(\beta_1,\beta_2,\alpha_1,\dots,\alpha_k)=
\left(-\frac{1}{4}\right)^{\frac{k-1}{2}}\,
\exp\left\{\sum_{i=2}^k\alpha_i-\frac{k-1}{2}(\beta_1+\beta_2)\right\}
Q_k(\alpha_1,\dots,\alpha_k)\,,
\label{oe2}
\end{equation}
where
\be
Q_{k+2}(\beta_1-\Delta,\beta_2-\Delta,\alpha_1,\dots,\alpha_k)
\cong\exp\left\{
(k-1)\Delta\right\}\,
N_k(\beta_1,\beta_2,\alpha_1,\dots,\alpha_k)\,.
\end{equation}
We have checked (\ref{oe2}) for $k=3,5$.

\subsection{Even--even clustering}

In this case the leading order reduced form factor clustering is of
the form
\be
\begin{split}
F^{xy}_{a_1\dots a_kb_1\dots b_l}
&(\alpha_1+\Delta,\dots,\alpha_k+\Delta,\beta_1,\dots,\beta_l)\\
&\qquad\cong \left(\frac{2\pi}{n\Delta}\right)
\left(\frac{\rme^{\Delta/2}}{2}\right)^{kl}
\exp\left\{\frac{l}{2}\sum_{i=1}^k\alpha_i
-\frac{k}{2}\sum_{j=1}^l\beta_j\right\}\\
&\Big\{F^{xq}_{a_1\dots a_k}(\alpha)
F^{yq}_{b_1\dots b_l}(\beta)
-F^{yq}_{a_1\dots a_k}(\alpha)
F^{xq}_{b_1\dots b_l}(\beta)\\
&\qquad +\tilde F^{xq}_{a_1\dots a_k}(\alpha)
\tilde F^{yq}_{b_1\dots b_l}(\beta)
-\tilde F^{yq}_{a_1\dots a_k}(\alpha)
\tilde F^{xq}_{b_1\dots b_l}(\beta)\Big\}
\end{split}
\end{equation}
and
\be
\begin{split}
\tilde F^{xy}_{a_1\dots a_kb_1\dots b_l}
&(\alpha_1+\Delta,\dots,\alpha_k+\Delta,\beta_1,\dots,\beta_l)\\
&\qquad\cong -\left(\frac{2\pi}{n\Delta}\right)
\left(\frac{\rme^{\Delta/2}}{2}\right)^{kl}
\exp\left\{\frac{l}{2}\sum_{i=1}^k\alpha_i
-\frac{k}{2}\sum_{j=1}^l\beta_j\right\}\\
&\Big\{F^{xq}_{a_1\dots a_k}(\alpha)
\tilde F^{yq}_{b_1\dots b_l}(\beta)
+F^{yq}_{a_1\dots a_k}(\alpha)
\tilde F^{xq}_{b_1\dots b_l}(\beta)\\
&\qquad -\tilde F^{xq}_{a_1\dots a_k}(\alpha)
F^{yq}_{b_1\dots b_l}(\beta)
-\tilde F^{yq}_{a_1\dots a_k}(\alpha)
F^{xq}_{b_1\dots b_l}(\beta)\\
&\qquad +2\tilde F^{xq}_{a_1\dots a_k}(\alpha)
\tilde F^{yq}_{b_1\dots b_l}(\beta)
+2\tilde F^{yq}_{a_1\dots a_k}(\alpha)
\tilde F^{xq}_{b_1\dots b_l}(\beta)\Big\}\,.
\end{split}
\end{equation}
Using the asymptotic relations
\be
\begin{split}
P^{(c)}_{l+2}(\alpha_1+\Delta,\beta_1,\alpha_2+\Delta,\beta_2,
\dots,\beta_l)&\cong\exp\left\{(l-2)\Delta\right\}
X_l(\alpha_1,\alpha_2,\beta_1,\dots,\beta_l)\\
P^{(t)}_{l+2}(\alpha_1+\Delta,\beta_1,\alpha_2+\Delta,\beta_2,
\dots,\beta_l)&\cong\exp\left\{\left(l-\frac{3}{2}\right)\Delta\right\}
Y_l(\alpha_1,\alpha_2,\beta_1,\dots,\beta_l)\\
\end{split}
\end{equation}
we have established the recursion relations
\be
\begin{split}
X_l&(\alpha_1,\alpha_2,\beta_1,\dots,\beta_l)=
\left(-\frac{1}{4}\right)^{\frac{l-2}{2}}\exp\left\{
\frac{l-2}{2}(\alpha_1+\alpha_2)-\sum_{j=3}^l\beta_j\right\}\\
&2\cosh\left(\frac{\beta_1-\beta_2}{2}\right)\Bigg\{
\cosh\left(\frac{\alpha_1-\alpha_2}{2}\right)P^{(t)}_l(\beta_1,\dots,
\beta_l)\\
&\qquad\qquad-\sinh\left(\frac{\alpha_1-\alpha_2}{2}\right)
\sinh\left(\frac{\beta_1-\beta_2}{2}\right)
P^{(c)}_l(\beta_1,\dots,\beta_l)\Bigg\}
\end{split}
\end{equation}
and
\be
\begin{split}
&\rme^{\frac{1}{2}(\beta_1-\alpha_1)}
Y_l(\alpha_1,\alpha_2,\beta_1,\dots,\beta_l)=
\left(-\frac{1}{4}\right)^{\frac{l-2}{2}}\exp\left\{
\frac{l-2}{2}(\alpha_1+\alpha_2)-\sum_{j=3}^l\beta_j\right\}\\
&\cosh\left(\frac{\beta_1-\beta_2}{2}\right)\Bigg\{
\left[\sinh\left(\frac{\alpha_1-\alpha_2}{2}\right)
-2\cosh\left(\frac{\alpha_1-\alpha_2}{2}\right)\right]
P^{(t)}_l(\beta_1,\dots,\beta_l)\\
&\qquad\qquad+\cosh\left(\frac{\alpha_1-\alpha_2}{2}\right)
\sinh\left(\frac{\beta_1-\beta_2}{2}\right)
P^{(c)}_l(\beta_1,\dots,\beta_l)\Bigg\}\,.
\end{split}
\end{equation}
We have verified the above relations for $l=2,4$.

%\vfill
%\eject

\section{Clustering in the O(3), O(4) models}

\subsection{$n=3$}

We recall the discussion in \cite{BN}. One has for the reduced form 
factors (see Appendix~B):
\ba
&&g^a_{b_1\dots b_ma_1\dots a_k}(\beta_1,\dots,\beta_m,\alpha_1+\triangle,
\dots,\alpha_k+\triangle)
\nonumber
\\
&&=\triangle^{km-1}\epsilon_{abc}
g^b_{a_1\dots a_k}(\alpha_1,\dots,\alpha_k)
g^c_{b_1\dots b_m}(\beta_1,\dots,\beta_m)+\rmO(\triangle^{km-2})\,,
\label{cluster1}
\ea
and similarly 
\ba
&&h_{b_1\dots b_ma_1\dots a_k}(\beta_1,\dots,\beta_m,\alpha_1+\triangle,
\dots,\alpha_k+\triangle)
\nonumber
\\
&&=\triangle^{km-2}
g^a_{a_1\dots a_k}(\alpha_1,\dots,\alpha_k)
g^a_{b_1\dots b_m}(\beta_1,\dots,\beta_m)+\rmO(\triangle^{km-3})\,.
\label{cluster0}
\ea                      
Note that in (\ref{cluster1}) members of the isospin 1 family are mapped
into themselves, while in (\ref{cluster0}) members of the isospin 1 family
are linked to members of the isospin 0 family. Observe also
that there is no distinction between the factorization properties
between even and odd members of the same family.

For the special case of $k=1\,,m=r-1$ the clustering relations read
\ba
g^a_{a_1\dots a_r}(\beta_1,\dots,\beta_r)&=&
\beta_r^{r-2}\epsilon_{aa_rb}
g^b_{a_1\dots a_{r-1}}(\beta_1,\dots,\beta_{r-1})+\rmO(\beta_r^{r-3})\,,
\\
h_{a_1\dots a_r}(\beta_1,\dots,\beta_r)&=&
\beta_r^{r-3}
g^{a_r}_{a_1\dots a_{r-1}}(\beta_1,\dots,\beta_{r-1})+\rmO(\beta_r^{r-4})\,,
\ea
which is in accordance with the propery that the reduced form factors 
are polynomials
of partial degree $(r-2)$ and $(r-3)$ in the isospin 1 and 0 case,
respectively. Since the product $\Psi_r$ also factorizes under clustering, 
the full (scalarized) form factors also satisfy clustering relations,
which are similar to (\ref{cluster1}) and (\ref{cluster0}).
For the $l=1$ family they can be found in Smirnov's book \cite{Smirnov}.

The clustering relations closely resemble some classical equations 
satisfied by the operators. For example dividing an even number of
particles into two odd clusters, (\ref{cluster1}) can be interpreted
as the quantum counterpart of the current in terms of the spin
operators. The division of an even number of particles into
two even clusters on the other hand, resembles the classical equation
$\partial_{\mu}J^a_{\nu}-\partial_{\nu}J^a_{\mu}\propto
\epsilon_{abc}J^b_{\mu}J^c_{\nu}\,.$
Finally the clustering of an odd number of particles corresponds to
$\partial_{\mu}s^a\propto\epsilon_{abc}s^bJ^c_{\mu}$.
Similarly (\ref{cluster0}) corresponds to the defining equation
for the energy momentum tensor in terms of the spin fields
or equivalently to its Sugawara form $T_{\mu\nu}\propto
J_{\mu}^aJ_{\nu}^a-\frac12\eta_{\mu\nu}J_{\rho}^aJ_{\rho}^a$.

In ref.~\cite{BN} the clustering properties above were used 
in particular to deduce the clustering properties of absolute
squares of the form factors, summed over internal symmetry indices
which enter in the expressions for spectral densities.

\subsection{O$(4)$ form factors: a three--particle example}

Just as the O$(4)$ S-matrix (\ref{minSS}) is (minus) the tensor product 
of two chiral
Gross--Neveu S--matrices, the O$(4)$ form factors can be written as 
tensor products of two chiral Gross--Neveu form factors. More precisely, 
the O$(4)$ form factors can be written as linear combinations of several
such tensor products. This solution of the O$(4)$ form factor equations,
for the case of even particle numbers, was given by 
F.~A.~Smirnov~\cite{Smi2}. The odd particle form factors must have a
similar structure. The solution for the three--particle form factors
of the O$(4)$ field operator was found by M.~Karowski~\cite{Michael}:
\begin{equation}
%\begin{split}
f_{P;ABC}(\theta_1,\theta_2,\theta_3)=D(\theta_1,\theta_2,\theta_3)
\sum_\omega\,
\widetilde F^{(\omega)}_{p_1;a_1b_1c_1}(\theta_1,\theta_2,\theta_3)\,
\widetilde F^{(\overline\omega)}_{p_2;a_2b_2c_2}
(\theta_1,\theta_2,\theta_3)\,.
%\end{split}
\label{Dtensor}
\end{equation}
Here $\widetilde F^{(\omega)}_{p;abc}(\theta_1,\theta_2,\theta_3)$
for $\omega=\pm$ are the spin $s=\pm \frac{1}{4}$ SU$(2)$--symmetric 
chiral Gross--Neveu model form factors discussed in Appendix~D. 
They satisfy the following homogeneous bootstrap equations: 
\begin{eqnarray}
\widetilde F^{(\omega)}_{p;i_1i_2i_3}(\theta_1+\lambda,\theta_2
+\lambda,\theta_3+\lambda)&=&
{\rm e}^{\omega\lambda/4}\,\widetilde F^{(\omega)}_{p;i_1i_2i_3}
(\theta_1,\theta_2,\theta_3)\,,
\label{tilFFi}\\
\widetilde F^{(\omega)}_{p;i_1i_2i_3}(\theta_1,\theta_2,\theta_3)&=&
\widetilde S^{uv}_{i_2i_3}(\theta_2-\theta_3)\,
\widetilde F^{(\omega)}_{p;i_1vu}(\theta_1,\theta_3,\theta_2)\,,
\label{tilFFii}\\
\widetilde F^{(\omega)}_{p;i_1i_2i_3}(\theta_1+2\pi i,\theta_2,\theta_3)&=&
-i\omega\,\widetilde F^{(\omega)}_{p;i_2i_3i_1}(\theta_2,\theta_3,\theta_1)
\label{tilFFiii}
\end{eqnarray}
and the residue equations
\begin{equation}
\widetilde F^{(\omega)}_{p;i_1i_2i_3}(\alpha,\beta,\theta_3)\approx
\frac{-4}{\alpha-\beta-i\pi}\left\{
\widetilde c_{i_1i_2}\,\widetilde F^{(\omega)}_{p;i_3}(\theta_3)+
i\omega\,\widetilde c_{i_1k}\,\widetilde S^{kl}_{i_2i_3}(
\beta-\theta_3)\,\widetilde F^{(\omega)}_{p;l}(\theta_3)\right\}\,.
\label{tilFFiv}
\end{equation}
Here the chiral Gross--Neveu S-matrix $\widetilde S^{kl}_{ij}(\theta)$, 
the anti--symmetric charge conjugation matrix $\widetilde c_{i_1i_2}$
and the one-particle form factors $\widetilde F^{(\omega)}_{p;l}(\theta)$
are all defined in Appendix~D.

It is easy to show that the homogeneous form factor equations are satisfied
by (\ref{Dtensor}) if the scalar prefactor $D$ is shift--invariant, 
anti-symmetric under the exchange of any pair of rapidities and is 
$2\pi i$--periodic in all rapidity variables. It also has to have a first 
order zero at points where two rapidities differ by $i\pi$ in order to 
satisfy the residue equation as well. The solution is
\begin{equation}
D(\theta_1,\theta_2,\theta_3)=\frac{-i}{32}\prod_{i<j}\coth\left(
\frac{\theta_i-\theta_j}{2}\right)\,,
\end{equation}
which also has the right normalization. With this choice we have
\begin{equation}
\begin{split}
f_{P;ABC}(\alpha,\beta,\theta)\approx
\frac{i}{\alpha-\beta-i\pi}\sum_\omega
&\left\{\widetilde c_{a_1b_1}\,\widetilde c_{p_1c_1}+i\omega
\,\widetilde c_{a_1k_1}\,\widetilde c_{p_1l_1}\,\widetilde S^{k_1l_1}_
{b_1c_1}(\beta-\theta)\right\}\\
&\left\{\widetilde c_{a_2b_2}\,\widetilde c_{p_2c_2}+i\overline\omega
\,\widetilde c_{a_2k_2}\,\widetilde c_{p_2l_2}\,\widetilde S^{k_2l_2}_
{b_2c_2}(\beta-\theta)\right\}\\
=\frac{2i}{\alpha-\beta-i\pi}
&\Big\{\overline C_{AB}\,\,\overline C_{PC}\,-\,
\,\overline C_{AK}\,\,\overline C_{PL}\,\, S^{KL}_
{BC}(\beta-\theta)\Big\}\,.
\end{split}\nonumber
\end{equation}
Here the O$(4)$ charge conjugation matrix $\overline C_{AB}$ is defined
in (\ref{overC}).

We parameterize the O$(4)$ form factors as follows:
\begin{equation}
\begin{split}
f_{P;ABC}(\theta_1,\theta_2,\theta_3)&=
\overline C_{PA}\,\overline C_{BC}\, g_1(\theta_1,\theta_2,\theta_3)\\
&+\overline C_{PB}\,\overline C_{AC}\, g_2(\theta_1,\theta_2,\theta_3)+
\overline C_{PC}\,\overline C_{AB} \,g_3(\theta_1,\theta_2,\theta_3)\,.
\end{split}
\end{equation}
Using the tensor product solution we can write this as
\begin{equation}
\begin{split}
g_2+g_3&=2D\,F^{(+)}_-\,F^{(-)}_-\,,\\
g_2-g_3&=D\,\left(F^{(+)}_+\,F^{(-)}_- +
F^{(+)}_-\,F^{(-)}_+\right)\,,\\
4g_1+g_2+g_3&=2D\,F^{(+)}_+\,F^{(-)}_+\,.
\end{split}
\end{equation}
For $(\theta_1,\theta_2,\theta_3)=(\Delta,\alpha,\beta)$ and
using the large $\Delta$ expansion of the components $F^{(\omega)}_\pm$
described in Appendix~D we find for the O$(4)$ form factors
\begin{equation}
\begin{split}
g_2+g_3&\approx -\frac{2\pi}{\Delta^2}(i\pi-\xi)\psi_1(\xi)\left\{
g_0+\frac{1}{4}\left[
\Psi\left(\frac{1}{2}+\frac{\xi}{2\pi i}\right)+
\Psi\left(\frac{1}{2}-\frac{\xi}{2\pi i}\right)\right]
\right\}+\cdots,\\
g_2-g_3&\approx -\frac{2\pi\psi_1(\xi)}{\Delta-\frac{\alpha+\beta}{2}}
\left\{1+\frac{i\pi}{\Delta}+\cdots\right\},\\
4g_1+g_2+g_3&\approx\frac{3\pi}{2\Delta^2}\psi_1(\xi)(\xi+i\pi)+\cdots,
\end{split}
\nonumber
\end{equation}
where $\xi=\alpha-\beta$.
We see that this is exactly the same expansion as the one we found by 
directly solving the O$(4)$ form factor equations asymptotically
in Subsect.~4.5. 
Only the value of $g_0$ cannot be determined from the asymptotic 
solution. This we found in Appendix~D by expanding the complete 
solution for large $\Delta$:
\begin{equation}
g_0=1-\frac{1}{2}\,\ln\frac{8\Delta}{\pi}\,.
\end{equation}

%\vfill
%\eject

\section{Concluding remarks}

In the course of this work various intriguing relations
concerning form factor clustering in the \On sigma--models were discussed
and new structures revealed. The relationship of the pattern of
clustering to the classical field equations in the case of O(3)
has been previously known \cite{Smirnov,BN}. Some of these
patterns in particular those involving the Sugawara structure
of the energy momentum tensor and those involving the 
(non--Abelian) curl--freeness of the Noether current (which is
so important for integrability) probably extend to general $n$.
We have further formulated a conjecture in Subsect.~4.7 concerning the 
(on--shell) nature of clustering to the operator product expansion.

We have tested our ansatz in various examples. 
Firstly we checked that the solutions obtained
by solving the form factor equations in leading order $1/n$ coincided
with those obtained by the field theoretical approach to the model.
Although this is as generally expected, it constitutes yet another test 
of the proposed equivalence of the S--matrix bootstrap construction and
functional integral definition of the models. We also found
that the large $n$ limit and limit of large rapidity commute. 
Although this is observed in previous studies it is not an obvious fact
(recall that there are many examples where the large $n$ limit 
and limit of small rapidity do not commute).

The case $n=4$ is a special case. Here we studied in detail (to our 
knowledge for the first time) the 3--particle spin form factor. 
The tensor product structure is 
probably particular to this case but its general form involving
the hypergeometric functions may give a hint to the 
outstanding unsolved problem of the construction
of form factors for general $n>4$. Moreover in this case
we found that subleading terms in the form factor clustering 
involved also logarithms of the (large) rapidity shift.

As mentioned in the introduction it is not completely implausible
that some of the structural properties found concerning 
form factor clustering in integrable 2--d asymptotically free 
models have their analog in 4--dimensional models in particular
for processes when the kinematics effectively reduces the dimension
to 1+1. 

%\vfill
%\eject

\section*{Acknowledgments}

We are particularly indebted to Michael Karowski for
providing us with the solution~(\ref{Dtensor}).
We also acknowledge Simon Ruijsenaars for discussions. 
J.~B.~is grateful to the Max-Planck-Institut f\"ur Physik for its
hospitality. 
This investigation was supported in part by the Hungarian 
National Science Fund OTKA (under T049495 and T043159).

%\vfill
%\eject

\appendix
\renewcommand{\thesection}{Appendix~A. Explicit S--matrices}
\section{}
\renewcommand{\thesection}{A}

\subsection{The $n=3$ case}

For $n=3$ the kernel $\tilde K_3(\omega)={\rm e}^{-\pi\omega}$ and the 
integral in (\ref{sigma2}) is easily done. The result is well known 
\cite{ZZ}:
\begin{equation}
\begin{split}
\sigma_1(\theta)&=\frac{2\pi i\theta}{(\theta+i\pi)(\theta-2\pi i)},\\
\sigma_2(\theta)&=\frac{\theta(\theta-i\pi)}{(\theta+i\pi)(\theta-2\pi i)},\\
\sigma_3(\theta)&=\frac{2\pi i(i\pi-\theta)}{(\theta+i\pi)(\theta-2\pi i)}\,.
\end{split}
\end{equation}

\subsection{The $n=4$ case}

In this case (\ref{sigma13}) and (\ref{sigma2}) simplify to
\begin{eqnarray}
\sigma_1(\theta)&=&\frac{i\pi\theta}{(i\pi-\theta)^2}\,S^{(2)}(\theta)\,,
\nonumber\\
\sigma_2(\theta)&=&\frac{\theta}{\theta-i\pi}\,S^{(2)}(\theta)\,,
\qquad\qquad S^{(2)}(\theta)=-{\cal A}^2(\theta),
\\
\sigma_3(\theta)&=&\frac{i\pi}{i\pi-\theta}\,S^{(2)}(\theta)\,,
\nonumber\\
\nonumber
\end{eqnarray}
where 
\begin{equation}
{\cal A}(\theta)=-\exp\left\{
2i\int_0^\infty\,\frac{{\rm d}\omega}{\omega}\,
\frac{\sin(\theta\omega)}
{1+{\rm e}^{\pi\omega}}\right\}\,.
\label{calA}
\end{equation}

The O(4) S--matrix is here given in the real basis
\be
\vert a,\theta\rangle,\qquad\qquad a=1,2,3,4.
\end{equation}
It is useful to transform the particles into a complex SU$(2)\times
{\rm SU}(2)$ basis
\be
\vert A,\theta\rangle,\qquad\qquad A=++,--,+-,-+\,.
\end{equation}
The transformation and its inverse are given by
\be
\vert a,\theta\rangle=\Omega_a^{\phantom{a}A}\vert A,\theta\rangle\,,
\qquad\qquad
\vert A,\theta\rangle={\cal K}_A^{\phantom{A}a}\vert a,\theta\rangle\,,
\end{equation}
with ${\cal  
K}_A^{\phantom{A}a}\Omega_a^{\phantom{a}B}=\delta^B_A\,,\,\,\,
\Omega_a^{\phantom{a}A}{\cal  K}_A^{\phantom{A}b}=\delta^b_a$\,.

The transformation rule of the S--matrix is
\be
S^{CD}_{AB}(\theta)={\cal  K}_A^{\phantom{A}a}\,
{\cal K}_B^{\phantom{A}b}\,
\Omega_c^{\phantom{a}C}\,
\Omega_d^{\phantom{a}D}\,S^{cd}_{ab}(\theta).
\end{equation}
Using the O(4) S--matrix explicitly we get
\be
S_{AB}^{CD}(\theta)=\sigma_1(\theta)\,P^{CD}\,R_{AB}+
\sigma_2(\theta)\,\delta^C_A\,\delta^D_B+
\sigma_3(\theta)\,\delta^C_B\,\delta^D_A\,,
\label{SM}
\end{equation}
where
\be
P^{CD}=\Omega_x^{\phantom{a}C}\,
\Omega_x^{\phantom{a}D},\qquad\qquad
R_{AB}={\cal K}_A^{\phantom{A}x}\,
{\cal K}_B^{\phantom{A}x}\,.
\end{equation}

Next we define the SU$(2)\times {\rm SU}(2)$ basis explicitly. The O(4)
generators in the vector representation are
\be
\left(\tau^{ab}\right)_{xy}=i\left(\delta^{ax}\,\delta^{by}-
\delta^{ay}\,\delta^{bx}\right),\qquad\qquad
a,b=1,2,3,4\,.
\end{equation}
We now define
\be
V^k=\frac{1}{2}\,\epsilon^{klm}\,\tau^{lm}\qquad{\rm and}\qquad
A^k=\tau^{k4},
\qquad\qquad k,l,m=1,2,3\,.
\end{equation}
Further
\be
W^k_\pm=\frac{1}{2}\left(V^k\pm A^k\right)\,.
\end{equation}
These are the SU$(2)\times {\rm SU}(2)$ generators since
\be
\left[W^k_+,W^l_-\right]=0,\qquad\qquad
\left[W^k_\pm,W^l_\pm\right]=-i\epsilon^{klm}\,W^m_\pm\,.
\end{equation}
We now define SU$(2)\times {\rm SU}(2)$ particle states such that
\be
\vert ++,\theta\rangle\quad {\rm has\ eigenvalue\ }\quad\frac{1}{2}\quad
{\rm w.r.t}\quad W^3_+\quad{\rm and}\quad\frac{1}{2}\quad
{\rm w.r.t}\quad W^3_-\,,
\end{equation}
\be
\ \ \vert +-,\theta\rangle\quad {\rm has\ eigenvalue\ }\quad\frac{1}{2}\quad
{\rm w.r.t}\quad W^3_+\quad{\rm and}\quad-\frac{1}{2}\quad
{\rm w.r.t}\quad W^3_-\,,
\end{equation}
and so on. Here the phases are also important. We make the following choice
\begin{eqnarray}
\vert ++,\theta\rangle=\frac{1}{\sqrt{2}}\left\{
\vert 1,\theta\rangle-i\vert 2,\theta\rangle\right\}\,,\nonumber\\
\vert --,\theta\rangle=\frac{1}{\sqrt{2}}\left\{
\vert 1,\theta\rangle+i\vert 2,\theta\rangle\right\}\,,\nonumber\\
\vert +-,\theta\rangle=\frac{1}{\sqrt{2}}\left\{
i\vert 3,\theta\rangle+\vert 4,\theta\rangle\right\}\,,\nonumber\\
\vert -+,\theta\rangle=\frac{1}{\sqrt{2}}\left\{
i\vert 3,\theta\rangle-\vert 4,\theta\rangle\right\}\,.\nonumber\\
\nonumber
\end{eqnarray}
Now we calculate
\be
P^{CD}=\omega_C\,\delta^{C\bar D}\quad{\rm (no\ sum)},\qquad\quad
R_{AB}=\omega_A\,\delta_{A\bar B}\quad{\rm (no\ sum)},\qquad\quad
\end{equation}
where
\be
\omega_{++}=
\omega_{--}=1\qquad\qquad{\rm and}\qquad\qquad
\omega_{+-}=
\omega_{-+}=-1
\end{equation}
and $\bar B$ is the charge conjugate of $B$.

\subsection{Tensor product S--matrix}

The S--matrix of the SU$(2)$ chiral Gross--Neveu model is \cite{Smirnov}
\be
\widetilde S^{\gamma\delta}_{\alpha\beta}(\theta)=
\frac{{\cal A}(\theta)}{i\pi-\theta}\,\left\{
i\pi\,\delta^\gamma_\beta\,
\delta^\delta_\alpha\,-
\theta\,\delta^\gamma_\alpha\,
\delta^\delta_\beta\,\right\}\,,
\label{chiralGN}
\end{equation}
where $\alpha,\beta,\gamma,\delta=+,-$.
Taking the tensor product of two such S--matrices we get
\be
\begin{split}
-\widetilde S^{\gamma_1\delta_1}_{\alpha_1\beta_1}(\theta)\,
&\widetilde S^{\gamma_2\delta_2}_{\alpha_2\beta_2}(\theta)=
\sigma_2(\theta)\,
\delta^{\gamma_1}_{\alpha_1}\,\delta^{\gamma_2}_{\alpha_2}\,
\delta^{\delta_1}_{\beta_1}\,\delta^{\delta_2}_{\beta_2}\,
+\sigma_3(\theta)\,
\delta^{\gamma_1}_{\beta_1}\,\delta^{\gamma_2}_{\beta_2}\,
\delta^{\delta_1}_{\alpha_1}\,\delta^{\delta_2}_{\alpha_2}\,\\
&+\sigma_1(\theta)\,
\left(\delta^{\gamma_1}_{\beta_1}\,\delta^{\delta_1}_{\alpha_1}
-\delta^{\gamma_1}_{\alpha_1}\,\delta^{\delta_1}_{\beta_1}\right)\,
\left(\delta^{\gamma_2}_{\beta_2}\,\delta^{\delta_2}_{\alpha_2}
-\delta^{\gamma_2}_{\alpha_2}\,\delta^{\delta_2}_{\beta_2}\right)\,,
\end{split}
\label{minSS}
\end{equation}
which is the same as (\ref{SM}) if we make the
identification $A=(\alpha_1,\alpha_2)$ etc.

\vfill
\eject

\appendix
\renewcommand{\thesection}{Appendix~B. Some form factors for the case $n=3$}
\section{}
\renewcommand{\thesection}{B}

For $n=3$ we can define $J^a_\mu=\frac{1}{2}\epsilon^{abc}\,J^{bc}_\mu$
and rewrite (\ref{currFF}) as
\begin{equation}
\langle0\vert J^a_\mu(0)\vert b_1,\theta_1;\dots;b_r,\theta_r
\rangle^{{\rm in}}=
-i\epsilon_{\mu\alpha}q^\alpha\,f^a_{b_1\dots b_r}(\theta_1,\dots,\theta_r)\,.
\label{currFF3}
\end{equation}
Using this unified notation\footnote{Note that no confusion can arise
here since the \On spin field has non--vanishing form factors for odd 
number of particles only whereas the current form factors are 
non--vanishing for an even number of particles only.}
for the form factors of the O(3) field and the Noether current we can
introduce reduced form factors $g^a_{b_1\dots b_r}$ by
\begin{equation}
f^a_{b_1\dots b_r}(\theta_1,\dots,\theta_r)=
\Psi_r(\theta_1,\dots,\theta_r)\,
g^a_{b_1\dots b_r}(\theta_1,\dots,\theta_r)\,,
\end{equation}
where
\begin{equation}
\Psi_r(\theta_1,\dots,\theta_r)=\frac{1}{2}\,\pi^{\left(\frac{3r}{2}-1\right)}
\,\prod_{1\leq i<j\leq r}\,\psi(\theta_i-\theta_j)
\end{equation}
with
\begin{equation}
\psi(\theta)=\frac{\theta-i\pi}{\theta(2\pi i-\theta)}\,\tanh^2\left(
\frac{\theta}{2}\right)\,.
\end{equation}
Similarly for the energy--momentum tensor we define
\begin{equation}
f_{b_1\dots b_r}(\theta_1,\dots,\theta_r)=
\Psi_r(\theta_1,\dots,\theta_r)\,
g_{b_1\dots b_r}(\theta_1,\dots,\theta_r)\,.
\end{equation}
The advantage of using these reduced form factors is that they, with
only one exception, are polynomial expressions in the particle
rapidities \cite{BN}.

The first reduced form factors for the O$(3)$ field are $g^a_{b_1}(\theta_1)=
\delta^{ab_1}$ and 
\begin{equation}
\begin{split}
g^a_{b_1b_2b_3}(\theta_1,\theta_2,\theta_3)&=
\delta^{ab_3}\,\delta^{b_1b_2}\,(\theta_2-\theta_1)\\
&+\delta^{ab_2}\,\delta^{b_1b_3}\,(\theta_1-\theta_3-2\pi i)
+\delta^{ab_1}\,\delta^{b_2b_3}\,(\theta_3-\theta_2)\,.
\end{split}
\end{equation}
For the current we have
\begin{equation}
g^a_{b_1b_2}(\theta_1,\theta_2)=\epsilon^{ab_1b_2}
\end{equation}
and finally for the energy--momentum tensor\footnote{This is the only
exceptional, non--polynomial reduced form factor.}
\begin{equation}
g_{b_1b_2}(\theta_1,\theta_2)=\frac{\delta^{b_1b_2}}{\theta_1-\theta_2-i\pi}\,.
\end{equation}
Note that for $n=3$ the functions $\psi_0,\psi_1$ defined in
(\ref{2ff}) are given by
\begin{equation}
\psi_1(\theta)=-\frac{\pi^2}{2}\,\psi(\theta),\qquad\qquad
\psi_0(\theta)=\frac{i\pi^2\psi(\theta)}{\theta-i\pi}\,.
\end{equation}
Many further explicit examples can be found in ref.~\cite{BN}.

\vfill
\eject

\appendix
\renewcommand{\thesection}{Appendix~C. Current 6--particle function}
\section{}
\renewcommand{\thesection}{C}

\ba
&&P_6^{(c)}(\theta_1,\theta_2,\theta_3,\theta_4,\theta_5,\theta_6)
=\frac{1}{32}\Bigl[\Bigl\{
\nonumber\\
&&3\cosh\frac12\left(\theta_1+\theta_2+\theta_3-\theta_4-\theta_5-\theta_6\right)
+3\cosh\frac12\left(\theta_1+\theta_2+\theta_5-\theta_3-\theta_4-\theta_6\right)
\nonumber\\
&+&3\cosh\frac12\left(\theta_1+\theta_3+\theta_4-\theta_2-\theta_5-\theta_6\right)
+4\cosh\frac12\left(\theta_1+\theta_3+\theta_5-\theta_2-\theta_4-\theta_6\right)
\nonumber\\
&+&2\cosh\frac12\left(\theta_1+3\theta_3-\theta_2-\theta_4-\theta_5-\theta_6\right)
+2\cosh\frac12\left(\theta_3+3\theta_1-\theta_2-\theta_4-\theta_5-\theta_6\right)
\nonumber\\
&+&2\cosh\frac12\left(\theta_1+3\theta_5-\theta_2-\theta_3-\theta_4-\theta_6\right)
+2\cosh\frac12\left(\theta_5+3\theta_1-\theta_2-\theta_3-\theta_4-\theta_6\right)
\nonumber\\
&+&2\cosh\frac12\left(\theta_3+3\theta_5-\theta_1-\theta_2-\theta_4-\theta_6\right)
+2\cosh\frac12\left(\theta_5+3\theta_3-\theta_1-\theta_2-\theta_4-\theta_6\right)
\nonumber\\
&+&2\cosh\frac12\left(\theta_3+3\theta_4-\theta_1-\theta_2-\theta_5-\theta_6\right)
+2\cosh\frac12\left(\theta_5+3\theta_6-\theta_1-\theta_2-\theta_3-\theta_4\right)
\nonumber\\
&+&2\cosh\frac12\left(\theta_1+3\theta_2-\theta_3-\theta_4-\theta_5-\theta_6\right)
\nonumber\\
&+&2\cosh\frac12\left(\theta_1+\theta_2+3\theta_3-\theta_4-\theta_5-3\theta_6\right)
+2\cosh\frac12\left(\theta_1+\theta_2+3\theta_6-\theta_4-\theta_5-3\theta_3\right)
\nonumber\\
&+&
4\cosh\frac12\left(\theta_1+3\theta_2+\theta_3-\theta_4-\theta_5-3\theta_6\right)
+2\cosh\frac12\left(\theta_1+3\theta_2+\theta_3-\theta_5-\theta_6-3\theta_4\right)
\nonumber\\
&+&4\cosh\frac12\left(\theta_1+3\theta_2+\theta_5-\theta_3-\theta_6-3\theta_4\right)
+2\cosh\frac12\left(\theta_1+3\theta_2+\theta_5-\theta_3-\theta_4-3\theta_6\right)
\nonumber\\
&+&4\cosh\frac12\left(\theta_1+\theta_3+3\theta_4-\theta_2-\theta_5-3\theta_6\right)
+2\cosh\frac12\left(\theta_1+\theta_3+3\theta_4-3\theta_2-\theta_5-\theta_6\right)
\nonumber\\
&+&2\cosh\frac12\left(\theta_1+\theta_5+3\theta_6-3\theta_2-\theta_3-\theta_4\right)
\nonumber\\
&+&\cosh\frac12\left(3\theta_1+3\theta_2-\theta_3-\theta_4-\theta_5-3\theta_6\right)
+\cosh\frac12\left(3\theta_1+3\theta_2-\theta_3-\theta_5-\theta_6-3\theta_4\right)
\nonumber\\
&+&\cosh\frac12\left(3\theta_3+3\theta_4-\theta_1-\theta_2-\theta_5-3\theta_6\right)
+\cosh\frac12\left(3\theta_3+3\theta_4-\theta_1-\theta_5-\theta_6-3\theta_2\right)
\nonumber\\
&+&\cosh\frac12\left(3\theta_5+3\theta_6-\theta_1-\theta_2-\theta_3-3\theta_4\right)
+\cosh\frac12\left(3\theta_5+3\theta_6-\theta_1-\theta_3-\theta_4-3\theta_2\right)
\nonumber\\
&+&\cosh\frac12\left(\theta_1+3\theta_3+3\theta_4-\theta_2-3\theta_5-3\theta_6\right)
+\cosh\frac12\left(\theta_3+3\theta_1+3\theta_2-\theta_4-3\theta_5-3\theta_6\right)
\nonumber\\
&+&\cosh\frac12\left(\theta_5+3\theta_1+3\theta_2-\theta_6-3\theta_3-3\theta_4\right)\Bigr\}
\nonumber\\
&+&\Bigl\{\theta_3\leftrightarrow\theta_4)\Bigr\}
+\Bigl\{\theta_5\leftrightarrow\theta_6)\Bigr\}
+\Bigl\{\theta_3\leftrightarrow\theta_4\,,\theta_5\leftrightarrow\theta_6)\Bigr\}
\Bigr]+\Bigl[\theta_1\leftrightarrow\theta_2\Bigr]\,.
\ea

\vfill
\eject

\appendix
\renewcommand{\thesection}{Appendix~D. 
XY model and chiral Gross--Neveu model form factors}
\section{}
\renewcommand{\thesection}{D}

The bootstrap solution of the XY model \cite{XY} is based on the 
extremal Sine--Gordon S--matrix, which is given by
\begin{eqnarray}
S^{--}_{--}(\theta)&=&S^{++}_{++}(\theta)={\cal A}(\theta)\,,
%=-\exp\left\{
%2i\int_0^\infty\frac{{\rm d}\omega}{\omega}\,\frac{\sin\theta\omega}{
%{\rm e}^{\pi\omega}+1}\right\},
\\
S^{+-}_{+-}(\theta)&=&S^{-+}_{-+}(\theta)= 
\frac{\kappa\theta}{i\pi-\theta}\,{\cal A}(\theta)\,,
\\
S^{+-}_{-+}(\theta)&=&S^{-+}_{+-}(\theta)=
\frac{i\pi}{i\pi-\theta}\,{\cal A}(\theta)\,,
\end{eqnarray}
where ${\cal A}(\theta)$ is given in (\ref{calA}) 
and $\kappa=1$ for the XY model S--matrix. The choice $\kappa=-1$ gives
the SU$(2)$ symmetric chiral Gross--Neveu S--matrix (\ref{chiralGN}). 
We will keep the notation $S^{\gamma\delta}_{\alpha\beta}(\theta)$ for 
the XY model S--matrix and will denote the chiral Gross--Neveu S--matrix 
by $\widetilde S^{\gamma\delta}_{\alpha\beta}(\theta)$.

For the XY model the crossing relation is
\begin{equation}
S^{\gamma\delta}_{\alpha\beta}(i\pi-\theta)=
c_{\beta\mu}\,c_{\delta\nu}\,S^{\gamma\mu}_{\alpha\nu}(\theta)\,,
\end{equation}
where the charge conjugation matrix $c_{\alpha\beta}$ has 
non-vanishing components
\begin{equation}
c_{+-}=c_{-+}=1\,,
\end{equation}
whereas for the chiral Gross--Neveu case crossing is given by
\begin{equation}
\widetilde S^{\gamma\delta}_{\alpha\beta}(i\pi-\theta)=
\widetilde c_{\beta\mu}\,\widetilde c_{\delta\nu}\,
\widetilde S^{\gamma\mu}_{\alpha\nu}(\theta)\,,
\end{equation}
with
\begin{equation}
\widetilde c_{+-}=-\widetilde c_{-+}=i\,.
\end{equation}
As discussed in Appendix~A, the O$(4)$ S-matrix is (minus) the tensor 
product of two chiral Gross--Neveu S--matrices:
\begin{equation}
S^{CD}_{AB}(\theta)=-
\widetilde S^{\gamma_1\delta_1}_{\alpha_1\beta_1}(\theta)
\,\widetilde S^{\gamma_2\delta_2}_{\alpha_2\beta_2}(\theta)\,,
\end{equation}
where $A=(\alpha_1,\alpha_2)$ etc. The crossing relation is
\begin{equation}
S^{CD}_{AB}(i\pi-\theta)=
\overline C_{BM}\,\overline C_{DN}\,S^{CM}_{AN}(\theta)\,,
\end{equation}
with
\begin{equation}
\overline C_{AB}=\widetilde c_{\alpha_1\beta_1}
\,\widetilde c_{\alpha_2\beta_2}\,.
\label{overC}
\end{equation}

\subsection{SU$_{-1}(2)$ symmetry}

As is well known, the Sine--Gordon model has quantum group symmetry 
SU$_q(2)$, which becomes SU$_{-1}(2)$ in the extremal case. In this case 
the algebra
of generators $\tau_\pm$, $j$ is identical to the ordinary SU$(2)$ algebra:
\begin{equation}
[\tau_+,\tau_-]=2j,\qquad\qquad [j,\tau_\pm]=\pm\tau_\pm\,,
\end{equation}
it is only the co-product $\Delta$ that is different from the classical case:
\begin{equation}
\Delta(j)=j\otimes1+1\otimes j,\qquad\quad
\Delta(\tau_\pm)=\tau_\pm\otimes(-1)^{2j}+1\otimes\tau_\pm\,.
\end{equation}
This means that if we build tensor product representations from the
basic doublet representation, the representation matrices are given by
\begin{equation}
\Delta_2(\tau_\pm)=-\tau_\pm\otimes1+1\otimes\tau_\pm\,,
\end{equation}
\begin{equation}
\Delta_3(\tau_\pm)=\tau_\pm\otimes1\otimes1-1\otimes\tau_\pm\otimes1+
1\otimes1\otimes\tau_\pm
\end{equation}
etc. These are representation matrices of the classical SU$(2)$ algebra
and they are simply related to the usual ones. For the two--particle
case the relation is
\begin{equation}
\begin{split}
\vert ++\rangle_{\rm cl}&=\vert++\rangle\,,\\
\vert +-\rangle_{\rm cl}&=\vert+-\rangle\,,\\
\vert -+\rangle_{\rm cl}&=-\vert-+\rangle\,,\\
\vert --\rangle_{\rm cl}&=-\vert--\rangle\,,
\end{split}
\end{equation}
where the $\vert \alpha\beta\rangle_{\rm cl}$ states transform
according to the usual two--particle representation. Similarly for 
higher states we have
\begin{equation}
\vert \alpha_r\dots\alpha_2\alpha_1\rangle_{\rm cl}=
\prod_{l=1}^r\left(\alpha_l\right)^{l+1}
\vert\alpha_r\dots\alpha_2\alpha_1\rangle\,.
\label{mani}
\end{equation}

We denote the SU$(2)$ generators acting in the Hilbert space
by $\hat J$, $\hat T_\pm$. We will be looking for local fields $\phi_\pm(z)$
transforming as elements of an SU$(2)$ doublet: 
\begin{equation}
\begin{split}
2[\hat J,\phi_\pm(z)]=\pm\phi_\pm(z),\qquad\quad
&[\hat T_+,\phi_+(z)]=[\hat T_-,\phi_-(z)]=0\,,\\
[\hat T_+,\phi_-(z)]=&\phi_+(z),\qquad\qquad\quad
[\hat T_-,\phi_+(z)]=\phi_-(z)\,.
\end{split}
\end{equation}
Note that $(\phi_+)^\dagger\not=c\phi_-$ for any constant $c$ 
(the above equations cannot have such solutions) i.e. the doublet 
fields must be genuinely complex.

SU$(2)$ symmetry restricts the form factors of doublet operators.
For the 1--particle form factors we have
\begin{equation}
F_{p;a}(\theta)=\langle0\vert\phi_p(0)\vert a,\theta\rangle=iG(\theta)
\widetilde c_{pa}\,.
\end{equation}
For the 3-particle ones
\begin{equation}
F_{p;abc}(\theta_1,\theta_2,\theta_3)=
\langle0\vert\phi_p(0)\vert a,\theta_1;b,\theta_2;c,\theta_3\rangle
\end{equation}
we introduce the notation
\begin{equation}
\begin{split}
F_{-;++-}(\theta_1,\theta_2,\theta_3)&=F_1(\theta_1,\theta_2,\theta_3)\,,\\
F_{-;+-+}(\theta_1,\theta_2,\theta_3)&=F_2(\theta_1,\theta_2,\theta_3)\,,\\
F_{-;-++}(\theta_1,\theta_2,\theta_3)&=F_3(\theta_1,\theta_2,\theta_3)\,.
\end{split}
\label{F123}
\end{equation}
Note that all other components either vanish by charge conservation
or are related to these ones by charge conjugation:
\begin{equation}
F_{\bar p;\bar a\bar b\bar c}(\theta_1,\theta_2,\theta_3)=
-F_{p;abc}(\theta_1,\theta_2,\theta_3)\,.
\end{equation}
The restriction coming from SU$(2)$ symmetry is
\begin{equation}
F_1-F_2+F_3=0\,.
\label{symrel}
\end{equation}

We also introduce the form factors corresponding to the manifestly
SU$(2)$ invariant basis (\ref{mani}):
\begin{equation}
\widetilde F_{p;abc}(\theta_1,\theta_2,\theta_3)=
\langle0\vert\phi_p(0)\vert a,\theta_1;b,\theta_2;c,\theta_3\rangle_{\rm cl}\,.
\end{equation}
For these form factors we have
\begin{equation}
\widetilde F_1=F_1,\qquad\quad
\widetilde F_2=-F_2,\qquad\quad
\widetilde F_3=F_3\,,
\end{equation}
\begin{equation}
\widetilde F_{\bar p;\bar a\bar b\bar c}(\theta_1,\theta_2,\theta_3)=
\widetilde F_{p;abc}(\theta_1,\theta_2,\theta_3)
\end{equation}
and
\begin{equation}
\widetilde F_1+\widetilde F_2+\widetilde F_3=0\,.
\end{equation}

We note that the basic spin fields \cite{XY} of the XY model,
\begin{equation}
S^\pm(z)=S^1(z)\pm S^2(z)
\end{equation}
obviously satisfy $(S^+)^\dagger=S^-$ and hence cannot be elements of a 
doublet.

In the following we will consider the form factors of
not only the doublet operators but also more general, charge $-1$,
spin $s$ fields. We will use the notation (\ref{F123}) also for these
more general form factors. Of course, (\ref{symrel}) only holds for
the SU$(2)$ doublet case.

\subsection{3--particle form factor equations for general spin}

We recall the bootstrap equations satisfied by the form factors
of a charge $-1$, spin $s$ operator (which may or may not be the
lower component of an SU$(2)$ doublet). The homogenous equations are:
\begin{eqnarray}
F_{i_1i_2i_3}(\theta_1+\lambda,\theta_2+\lambda,\theta_3+\lambda)&=&
{\rm e}^{s\lambda}\,F_{i_1i_2i_3}(\theta_1,\theta_2,\theta_3),
\label{FFi}\\
F_{i_1i_2i_3}(\theta_1,\theta_2,\theta_3)&=&
S^{uv}_{i_2i_3}(\theta_2-\theta_3)\,
F_{i_1vu}(\theta_1,\theta_3,\theta_2),\label{FFii}\\
F_{i_1i_2i_3}(\theta_1+2\pi i,\theta_2,\theta_3)&=&
\eta_{i_1}\,F_{i_2i_3i_1}(\theta_2,\theta_3,\theta_1)\,.\label{FFiii}
\end{eqnarray}
This is supplemented by the residue equation
\begin{equation}
F_{i_1i_2i_3}(\alpha,\beta,\theta_3)\approx
\frac{4i}{\alpha-\beta-i\pi}\left\{
c_{i_1i_2}\,F_{i_3}(\theta_3)-\eta_{i_1}\,c_{i_1k}\,S^{kl}_{i_2i_3}(
\beta-\theta_3)\,F_l(\theta_3)\right\}\,.
\label{FFiv}
\end{equation}
Here and in the cyclic equation (\ref{FFiii}) $\eta_{i_1}$ is a phase factor
that expresses the relative non--locality between the field, whose form
factors we are constructing and the basic spin fields $S^\pm$ that create
the asymptotic particles using the LSZ asymptotic formula. Consistency 
between the cyclic equation (\ref{FFiii}) and the 
shift equation (\ref{FFi}) requires that
\begin{equation}
\eta_+=\eta={\rm e}^{2\pi is},\qquad\qquad \eta_-={\rm e}^{-2\pi is}
\end{equation}
and this is sufficient to determine the one--point function
\begin{equation}
F_j(\theta)=g\delta^+_j\,{\rm e}^{s\theta}
\end{equation}
up to the normalization constant $g$.

If we write the shift and cyclic equations in terms of the independent
components $F_1$, $F_2$, $F_3$ we get 
\begin{equation}
F_k(\theta_1+\lambda,\theta_2+\lambda,\theta_3+\lambda)=
{\rm e}^{s\lambda}\,F_k(\theta_1,\theta_2,\theta_3),\qquad k=1,2,3
\end{equation}
and
\begin{equation}
F_k(\theta_1+2\pi i,\theta_2,\theta_3)=
\eta\,F_{k+1}(\theta_2,\theta_3,\theta_1),\qquad k=1,2\,.
\label{FFcycl}
\end{equation}
The $k=3$ equation $F_3(\theta_1+2\pi i,\theta_2,\theta_3)=
\eta^{-1}\,F_1(\theta_2,\theta_3,\theta_1)$
is already a consequence of the above two.

We have seen that SU$(2)$ symmetry requires
\begin{equation}
\zeta=F_1-F_2+F_3=0\,.
\label{zeta}
\end{equation}
For later purposes we introduce
\begin{equation}
F_\pm=F_1\pm F_2\,,
\label{Fpm}
\end{equation}
in terms of which (\ref{zeta}) can also be written as $F_3=-F_-$.
From (\ref{FFcycl}) it follows that
\begin{equation}
\zeta(\theta_1+2\pi i,\theta_2,\theta_3)=
-\eta\zeta(\theta_2,\theta_3,\theta_1)
+\left(\eta+\frac{1}{\eta}\right)\,F_1(\theta_2,\theta_3,\theta_1)
\end{equation}
thus an SU$(2)$ doublet field must have $\eta=\pm i$ i.e. spin 
$s=\pm1/4\,({\rm mod}\,1)$. We will consider two such doublet solutions 
$\phi^{(\omega)}_p(z)$ with $s=\omega/4,\quad\eta=\omega i$ ($\omega=\pm$).
It is natural to write the form factors in this case using the manifestly
symmetric basis vectors (\ref{mani}).
We choose the normalization ($g^{(\pm)}=2$) such that the 1-particle 
form factors are given by
\begin{equation}
\widetilde F^{(\omega)}_{p;a}(\theta)=
\langle0\vert\phi^{(\omega)}_p(0)\vert a,\theta\rangle
=2i\widetilde c_{pa}{\rm e}^{\omega\theta/4}\,.
\end{equation}

Written in terms of the form factors of these operators, the three--particle
equations (\ref{FFi}-\ref{FFiv}) become the form factor equations
(\ref{tilFFi}-\ref{tilFFiv}), discussed in the main text.

\subsection{Reduced form factors}

To simplify the solution of the 3--particle form factor equations we 
introduce a set of \lq\lq reduced" form factors $f_m$ ($m=1,2,3$) by 
writing \cite{XY}
\begin{equation}
F_m(\theta_1,\theta_2,\theta_3)=-2\pi^2{\cal N}Y(\theta_1,\theta_2,\theta_3)
{\rm e}^{s(\theta_1+\theta_2+\theta_3)}f_m(\theta_1,\theta_2,\theta_3)\,,
\end{equation}
where the prefactor $Y$ is
\begin{equation}
Y(\theta_1,\theta_2,\theta_3)=\prod_{i<j}\,y(\theta_i-\theta_j)
\end{equation}
with
\begin{equation}
y(\theta)=\sinh\left(\frac{\theta}{2}\right){\rm e}^{E(\theta)}\,,
\end{equation}
where
\begin{equation}
E(\theta)=\int_0^\infty\frac{{\rm d}\omega}{\omega}\,
\frac{\left[\cosh\omega(\pi+i\theta)-1\right]}{\sinh\pi\omega}\,\frac{1}
{(1+{\rm e}^{\pi\omega})}\,.
\end{equation}
Finally the normalization constant is
\begin{equation}
{\cal N}=\frac{i}{\pi^{11/2}}{\rm e}^{-E(0)}{\rm e}^{-i\pi s}\,.
\end{equation}
Note that the function $E(\theta)$ is related to $\psi_1(\theta)$ 
(defined in (\ref{psi120})) for the case $n=4$ by
\begin{equation}
\psi_1(\theta)|_{n=4}=\frac{2i\sinh\left(\frac{\theta}{2}\right)}
{i\pi -\theta}\rme^{2E(\theta)}\,,
\end{equation}
and for large $\theta$ it behaves as
\begin{equation}
4E(\theta)=-\theta+\ln(2\theta)+2E(0)+i\pi \left(1-\theta^{-1}\right)
+\rmO(\theta^{-2})\,.
\end{equation}
For later use we introduce the function 
\begin{equation}
\Phi(\theta)=\Gamma\left(\frac{1}{2}+\frac{\theta}{2\pi i}\right)\,
\Gamma\left(-\frac{\theta}{2\pi i}\right)\,.
\end{equation}
With the help of this function we can write the S--matrix element 
${\cal A}(\theta)$ as
\begin{equation}
{\cal A}(\theta)=\frac{\Phi(\theta)}{\Phi(-\theta)}\,.
\end{equation}

For completeness, we give here the form factor equations, rewritten in terms
of the reduced form factors $f_m$:
\begin{equation}
f_m(\theta_1+\lambda,\theta_2+\lambda,\theta_3+\lambda)=
{\rm e}^{-2s\lambda}\,f_m(\theta_1,\theta_2,\theta_3)\,,
\end{equation}
\begin{equation}
\begin{split}
f_3(\alpha,\theta,\theta^\prime)&=f_3(\alpha,\theta^\prime,\theta)\,,\\
f_-(\alpha,\theta,\theta^\prime)&=f_-(\alpha,\theta^\prime,\theta)\,,\\
f_+(\alpha,\theta,\theta^\prime)&=\frac
{i\pi+\theta-\theta^\prime}{i\pi-\theta+\theta^\prime}\,
f_+(\alpha,\theta^\prime,\theta)\,,
\end{split}
\qquad f_\pm=f_1\pm f_2\,,
\end{equation}
\begin{equation}
f_m(\theta_1+2\pi i,\theta_2,\theta_3)=
f_{m+1}(\theta_2,\theta_3,\theta_1),\qquad (m=1,2)\,,
\end{equation}
\begin{equation}
\begin{split}
f_1(\alpha,\beta,\theta)&\approx \frac{-2ig\eta\pi^2{\rm e}^{-2s\beta}}
{\alpha-\beta-i\pi}\,\left\{
\frac{-i\pi}{i\pi-\beta+\theta}\,\Phi(\beta-\theta)\right\}\,,\\
f_2(\alpha,\beta,\theta)&\approx \frac{-2ig\pi^2{\rm e}^{-2s\beta}}
{\alpha-\beta-i\pi}\,\left\{\Phi(\theta-\beta)-
\frac{\eta(\beta-\theta)}{i\pi-\beta+\theta}\,\Phi(\beta-\theta)\right\}\,,\\
f_3(\alpha,\beta,\theta)&\approx \frac{-2ig\pi^2{\rm e}^{-2s\beta}}
{\alpha-\beta-i\pi}\,\left\{\Phi(\theta-\beta)-
\frac{1}{\eta}\,\Phi(\beta-\theta)\right\}\,.
\end{split}
\end{equation}

Later we will explicitly solve the form factor equations for the reduced
form factors and calculate their large rapidity limit. But even before 
having 
the complete solution, a lot of information about their large rapidity 
behavior can already be obtained by expanding the equations themselves.
We take the Ansatz
\begin{equation}
f_m(\Delta,\alpha,\beta)\approx\frac{{\rm e}^{-\frac{1}{2}\Delta}
{\rm e}^{k\alpha}}{(\Delta-\alpha)^p}\,\left\{
U_m(\alpha-\beta)+\frac{W_m(\alpha-\beta)}{\Delta-\alpha}+\cdots\right\}
\qquad (m=1,2,3)
\label{noLOG}
\end{equation}
for large $\Delta$, where $k=\frac{1}{2}-2s$. We get restrictions on the
leading power $p$ and the expansion coefficients $U_m$, $W_m$ by 
substituting this Ansatz into the reduced form factor equations. 
In particular, from the residue equations, using the asymptotic expansion 
of $\Phi$, we get for large $\Delta$
\begin{equation}
%\begin{split}
f_m(\alpha,\beta,\Delta)\approx\frac{-ic}{\alpha-\beta-i\pi}\,
{\rm e}^{k\beta}{\rm e}^{-\frac{1}{2}\Delta}
\left\{\begin{split}
&\frac{\eta}{(\Delta-\beta)^{3/2}}-\frac{3i\pi\eta}{4\Delta^{5/2}}+\cdots\\
&\frac{1+i\eta}{\pi}\,
\frac{1}{(\Delta-\beta)^{1/2}}+\frac{3\eta-i}{4}\,
\frac{1}{\Delta^{3/2}}+\cdots\\
&\frac{1}{\pi}\left(1-\frac{i}{\eta}\right)
\frac{1}{(\Delta-\beta)^{1/2}}+\frac{1}{4}\left(\frac{1}{\eta}-i\right)
\frac{1}{\Delta^{3/2}}+\cdots
\end{split}\right.
%\end{split}
\end{equation}
where
\begin{equation}
c=4\pi^4\sqrt{2\pi}{\rm e}^{-\frac{i\pi}{4}}g\,.
\end{equation}
From here we see that for most spin values for which $\eta\not=i$
($-\frac{1}{4}\leq s<\frac{1}{4}$), the leading power is $p=1/2$ but 
for $s=\frac{1}{4}$ we have $\eta=i$ and the leading power is $p=3/2$. 
It is easy to solve the form factor equations in this expanded form.
For the case $\eta=-i$ ($s=-\frac{1}{4}$) we get 
\begin{equation}
U_m(\xi)=\left(\begin{matrix}1\\1\\0\end{matrix}\right)\,U(\xi)\,,\qquad
\qquad W_3(\xi)=-W_-(\xi)\,,
\label{sol1}
\end{equation}
where
\begin{equation}
\begin{split}
U(\xi)&=-\frac{2c}{\pi}\,
\frac{{\rm e}^{-\frac{1}{2}\xi}}{(\pi i-\xi)}\,,\\
W_+(\xi)&=\frac{c}{\pi}\,\frac{{\rm e}^{-\frac{1}{2}\xi}}{(\pi i-\xi)}\,
(\xi-2\pi i),\\
W_-(\xi)&=-\frac{2c}{\pi}\,{\rm e}^{-\frac{1}{2}\xi}\,\left\{
g_0+\frac{1}{4}\left[
\Psi\left(\frac{1}{2}+\frac{\xi}{2\pi i}\right)+
\Psi\left(\frac{1}{2}-\frac{\xi}{2\pi i}\right)\right]\right\}\,,
\end{split}
\end{equation}
whereas for the case $\eta=i$ ($s=\frac{1}{4}$) we get 
\begin{equation}
U_m(\xi)=\left(\begin{matrix}-1/2\\1/2\\1\end{matrix}\right)\,U(\xi)\,,\qquad
\qquad W_3(\xi)=-W_-(\xi)
\end{equation}
and 
\begin{equation}
\begin{split}
U(\xi)&=\frac{ic}{2\cosh\frac{\xi}{2}}\,,\\
W_-(\xi)&=\frac{3ic}{8}\,\frac{(\xi-2\pi i)}
{\cosh\frac{\xi}{2}},\\
W_+(\xi)&=\frac{3ic}{16}\,
\frac{(\xi+\pi i)}{\cosh\frac{\xi}{2}}\,.
\end{split}
\label{sol4}
\end{equation}

As discussed in Section~4, there is a logarithmic piece in the
subleading term of the O$(4)$ form factors. As we will see later, such
terms are also present in the asymptotic expansion of the spin~$\pm1/4$
form factors. Anticipating this fact we extend (\ref{noLOG}) by a 
logarithmic piece of the form
\begin{equation}
%f_m(\Delta,\alpha,\beta)\approx
\frac{{\rm e}^{-\frac{1}{2}\Delta}
{\rm e}^{k\alpha}}{(\Delta-\alpha)^{\widetilde p}}\,\ln(\Delta-\alpha)\,
\left\{\widetilde U_m(\alpha-\beta)+\frac{
\widetilde W_m(\alpha-\beta)}{\Delta-\alpha}+\cdots\right\}\,.
%\qquad (m=1,2,3)
\label{LOG}
\end{equation}
The functions $\widetilde U_m$ and $\widetilde W_m$ satisfy very
similar equations to the ones discussed above for $U_m$ and $W_m$. The
main difference is that the residue equations are free of logs. We
find that non--vanishing solutions are only possible for 
$\widetilde p=1/2$ or $\widetilde p=3/2$. Finally requiring regularity 
(also at infinity) eliminates all but one possibility: $\widetilde
p=3/2$ for $s=-1/4$ with solution
\begin{equation}
\widetilde U^{(-)}_-(\xi)=\ell_0\,{\rm e}^{-\frac{1}{2}\xi}\,,
\end{equation}
with arbitrary constant $\ell_0$. This means that the full asymptotic
expansion (the sum of (\ref{noLOG}) and (\ref{LOG})) is correctly
given by (\ref{noLOG}) alone, with solution (\ref{sol1}-\ref{sol4}),
provided only that we allow the \lq\lq constant" $g_0$ depend
(linearly) on $\ln\Delta$. The actual value of $g_0$ can be 
calculated from the large $\Delta$ expansion of the exact solution.
We now turn to this calculation.

\subsection{Contour integral solution}

H. Babujian et al. \cite{BFKZ} found the solution of the reduced form 
factor equations in terms of a contour integral:
\begin{equation}
f_m(\theta_1,\theta_2,\theta_3)=-\frac{1}{2\pi^2}\,\int_{\cal C}\,
{\rm d}u\,{\rm e}^{-2su}\,
t_m(\theta_1,\theta_2,\theta_3;u)\prod_{j=1}^3\Phi(\theta_j-u)\,,
\end{equation}
where
\begin{equation}
\begin{split}
t_1(\theta_1,\theta_2,\theta_3;u)&=
\frac{\theta_1-u}{i\pi-\theta_1+u}\,
\frac{\theta_2-u}{i\pi-\theta_2+u}\,
\frac{i\pi}{i\pi-\theta_3+u}\,,\\
t_2(\theta_1,\theta_2,\theta_3;u)&=
\frac{\theta_1-u}{i\pi-\theta_1+u}\,
\frac{i\pi}{i\pi-\theta_2+u}\,,\\
t_3(\theta_1,\theta_2,\theta_3;u)&=
\frac{i\pi}{i\pi-\theta_1+u}
\end{split}
\end{equation}
and the contour ${\cal C}$ (for real rapidities $\theta_i$) comes from
$-\infty$ along a line parallel to the real axis and going somewhat below the
singular points $\theta_i-i\pi$, then turns back and goes around the points
$\theta_i$ before it goes to $+\infty$ again parallel to the real line.
Precisely this integral along such a contour is the special function
known as Meijer's G--function \cite{Wolfram}:
\begin{equation}
f_m(\theta_1,\theta_2,\theta_3)=G^{33}_{33}\left(
{\rm e}^{-4\pi is}\Bigg\vert\begin{matrix}
a_1^{(m)}&a_2^{(m)}&a_3^{(m)}\\
b_1^{(m)}&b_2^{(m)}&b_3^{(m)}\end{matrix}\right)\,.
\end{equation}
The parameters depend on the rapidities:
\begin{equation}
\begin{split}
&{
\begin{split}
a_1^{(1)}&=-\frac{i\theta_1}{2\pi}\,,\\
b_1^{(1)}&=-\frac{1}{2}-\frac{i\theta_1}{2\pi},
\end{split}\qquad\qquad
\begin{split}
a_2^{(1)}&=-\frac{i\theta_2}{2\pi}\,,\\
b_2^{(1)}&=-\frac{1}{2}-\frac{i\theta_2}{2\pi}\,,
\end{split}\qquad\qquad
\begin{split}
a_3^{(1)}&=1-\frac{i\theta_3}{2\pi}\,,\\
b_3^{(1)}&=-\frac{1}{2}-\frac{i\theta_3}{2\pi}\,,
\end{split}
}\\
&{
\begin{split}
a_1^{(2)}&=-\frac{i\theta_1}{2\pi}\,,\\
b_1^{(2)}&=-\frac{1}{2}-\frac{i\theta_1}{2\pi}\,,
\end{split}\qquad\qquad
\begin{split}
a_2^{(2)}&=1-\frac{i\theta_2}{2\pi}\,,\\
b_2^{(2)}&=-\frac{1}{2}-\frac{i\theta_2}{2\pi}\,,
\end{split}\qquad\qquad
\begin{split}
a_3^{(2)}&=1-\frac{i\theta_3}{2\pi}\,,\\
b_3^{(2)}&=\frac{1}{2}-\frac{i\theta_3}{2\pi}\,,
\end{split}
}\\
&{
\begin{split}
a_1^{(3)}&=1-\frac{i\theta_1}{2\pi}\,,\\
b_1^{(3)}&=-\frac{1}{2}-\frac{i\theta_1}{2\pi}\,,
\end{split}\qquad\qquad
\begin{split}
a_2^{(3)}&=1-\frac{i\theta_2}{2\pi}\,,\\
b_2^{(3)}&=\frac{1}{2}-\frac{i\theta_2}{2\pi}\,,
\end{split}\qquad\qquad\quad
\begin{split}
a_3^{(3)}&=1-\frac{i\theta_3}{2\pi}\,,\\
b_3^{(3)}&=\frac{1}{2}-\frac{i\theta_3}{2\pi}\,.
\end{split}
}
\end{split}
\end{equation}
Finally we note that Meijer's G--function $G^{33}_{33}$ can be expressed
in terms of Gamma functions and hypergeometric functions as 
follows \cite{Wolfram}:
\begin{equation}
\begin{split}
&G^{33}_{33}\left(z\Big\vert\begin{matrix}a_1&a_2&a_3\\b_1&b_2&b_3
\end{matrix}\right)\\&=
z^{b_1}\,\Omega(1-a_1+b_1,1-a_2+b_1,1-a_3+b_1;
b_2-b_1,b_3-b_1;-z)+2 {\rm \ perms}\,,
\end{split}
\end{equation}
where
\begin{equation}
%\begin{split}
\Omega(u_1,u_2,u_3;v_1,v_2;z)=
\Gamma(u_1)\Gamma(u_2)\Gamma(u_3)\Gamma(v_1)\Gamma(v_2)\,
_3F_2(u_1,u_2,u_3;1-v_1,1-v_2;z)\,.
%\end{split}
\end{equation}

This can be used to express the three--particle form factors as
follows.
\begin{equation}
\begin{split}
f_1(\theta_1,\theta_2,\theta_3)&=
{\rm e}^{2\pi is}\,{\rm e}^{-2s\theta_1}\,\Omega\left(
\frac{1}{2},
\frac{1}{2}-\frac{i\theta_{12}}{2\pi},
-\frac{1}{2}-\frac{i\theta_{13}}{2\pi};
\frac{i\theta_{12}}{2\pi},
\frac{i\theta_{13}}{2\pi};
-{\rm e}^{-4\pi is}\right)\\
&+{\rm e}^{2\pi is}\,{\rm e}^{-2s\theta_2}\,\Omega\left(
\frac{1}{2}+\frac{i\theta_{12}}{2\pi},
\frac{1}{2},
-\frac{1}{2}-\frac{i\theta_{23}}{2\pi};
-\frac{i\theta_{12}}{2\pi},
\frac{i\theta_{23}}{2\pi};
-{\rm e}^{-4\pi is}\right)\\
&+{\rm e}^{2\pi is}\,{\rm e}^{-2s\theta_3}\,\Omega\left(
\frac{1}{2}+\frac{i\theta_{13}}{2\pi},
\frac{1}{2}+\frac{i\theta_{23}}{2\pi},
-\frac{1}{2};
-\frac{i\theta_{13}}{2\pi},
-\frac{i\theta_{23}}{2\pi};
-{\rm e}^{-4\pi is}\right)\,,
\end{split}
\label{f1ex}
\end{equation}
\begin{equation}
\begin{split}
f_2(\theta_1,\theta_2,\theta_3)&=
{\rm e}^{2\pi is}\,{\rm e}^{-2s\theta_1}\,\Omega\left(
\frac{1}{2},
-\frac{1}{2}-\frac{i\theta_{12}}{2\pi},
-\frac{1}{2}-\frac{i\theta_{13}}{2\pi};
\frac{i\theta_{12}}{2\pi},
1+\frac{i\theta_{13}}{2\pi};
-{\rm e}^{-4\pi is}\right)\\
&+{\rm e}^{2\pi is}\,{\rm e}^{-2s\theta_2}\,\Omega\left(
\frac{1}{2}+\frac{i\theta_{12}}{2\pi},
-\frac{1}{2},
-\frac{1}{2}-\frac{i\theta_{23}}{2\pi};
-\frac{i\theta_{12}}{2\pi},
1+\frac{i\theta_{23}}{2\pi};
-{\rm e}^{-4\pi is}\right)\\
&+{\rm e}^{-2\pi is}\,{\rm e}^{-2s\theta_3}\,\Omega\left(
\frac{3}{2}+\frac{i\theta_{13}}{2\pi},
\frac{1}{2}+\frac{i\theta_{23}}{2\pi},
\frac{1}{2};
-1-\frac{i\theta_{13}}{2\pi},
-1-\frac{i\theta_{23}}{2\pi};
-{\rm e}^{-4\pi is}\right)\,,
\end{split}
\label{f2ex}
\end{equation}
\begin{equation}
\begin{split}
f_3(\theta_1,\theta_2,\theta_3)&=
{\rm e}^{2\pi is}\,{\rm e}^{-2s\theta_1}\,\Omega\left(
-\frac{1}{2},
-\frac{1}{2}-\frac{i\theta_{12}}{2\pi},
-\frac{1}{2}-\frac{i\theta_{13}}{2\pi};
1+\frac{i\theta_{12}}{2\pi},
1+\frac{i\theta_{13}}{2\pi};
-{\rm e}^{-4\pi is}\right)\\
&+{\rm e}^{-2\pi is}\,{\rm e}^{-2s\theta_2}\,\Omega\left(
\frac{1}{2}+\frac{i\theta_{12}}{2\pi},
\frac{1}{2},
\frac{1}{2}-\frac{i\theta_{23}}{2\pi};
-1-\frac{i\theta_{12}}{2\pi},
\frac{i\theta_{23}}{2\pi};
-{\rm e}^{-4\pi is}\right)\\
&+{\rm e}^{-2\pi is}\,{\rm e}^{-2s\theta_3}\,\Omega\left(
\frac{1}{2}+\frac{i\theta_{13}}{2\pi},
\frac{1}{2}+\frac{i\theta_{23}}{2\pi},
\frac{1}{2};
-1-\frac{i\theta_{13}}{2\pi},
-\frac{i\theta_{23}}{2\pi};
-{\rm e}^{-4\pi is}\right)\,.
\end{split}
\label{f3ex}
\end{equation}
We are interested in the asymptotics of these form factors in the
limit $\theta_1\to+\infty$. The exponential part of the form factor
asymptotics, which comes entirely from the Gamma functions, is
\begin{equation}
{\rm e}^{-(1+2s)\theta_1}
\end{equation}
for the first of the three terms
for all $f_m$ and is
\begin{equation}
{\rm e}^{-\frac{1}{2}\theta_1}
\label{half}
\end{equation}
for the second and third terms. Thus the exponential part of the 
asymptotics is given by (\ref{half}), which comes from the second and third
terms in almost all cases, except for $s=-1/4$, in which case
also the first terms contribute.

To calculate the leading asymptotics of our form factors we will need
the asymptotic behavior of the generalized hypergeometric functions
$_3F_2$ in the case of some of its parameters large. We will use
a simple integral representation \cite{Wolfram2} of this function 
to establish the asymptotic formulae we need in this calculation.
\begin{equation}
\begin{split}
&_3F_2(a_1,a_2,a_3;b_1,b_2;z)\\
&=\frac{\Gamma(b_2)}{\Gamma(a_3)\Gamma(b_2-a_3)}\,\int_0^1{\rm d}t\,
t^{a_3-1}\,(1-t)^{b_2-a_3-1}\,_2F_1(a_1,a_2;b_1;tz)\,,
\end{split}
\label{int32}
\end{equation}
which is valid for ${\rm Re}(b_2)>{\rm Re}(a_3)>0$ in the range
$\vert z\vert<1$, but can be extended to the limit $\vert z\vert\to1$.
The Gauss hypergeometric function $_2F_1$ can in turn be expressed as the
integral
\begin{equation}
%\begin{split}
_2F_1(a,b;c;z)
=\frac{\Gamma(c)}{\Gamma(b)\Gamma(c-b)}\,\int_0^1{\rm d}t\,
t^{b-1}\,(1-t)^{c-b-1}\,(1-tz)^{-a}\,,
%\end{split}
\label{int21}
\end{equation}
valid for ${\rm Re}(c)>{\rm Re}(b)>0$.

Using both integral representations above
simultaneously we can show that for large (real) $\lambda$
\begin{equation}
\begin{split}
_3F_2(a_1,\alpha_2+i\lambda,&\alpha_3+i\lambda;\beta_1+i\lambda,
\beta_2+i\lambda;1)\\
&\approx\frac{\Gamma(\beta_1+\beta_2-\alpha_2-\alpha_3-a_1)}
{\Gamma(\beta_1+\beta_2-\alpha_2-\alpha_3)}\,
\vert\lambda\vert^{a_1}\,{\rm e}^{\frac{i\pi a_1}{2}
{\rm sgn}\,\lambda}\,,
\end{split}
\label{F32as}
\end{equation}
valid for ${\rm Re}(\beta_2)>{\rm Re}(\alpha_3)>0$ and
${\rm Re}(\beta_1)>{\rm Re}(\alpha_2)>0$. We are sure however that
this estimate holds in a larger range of parameters. In particular
to get the contribution of the first terms for $f_1$, $f_2$
we need an estimate of
the lhs of (\ref{F32as})  
for $a_1=1/2,\alpha_2=\frac12\rho+\frac{i\theta_2}{2\pi},
\alpha_3=-\frac12+\frac{i\theta_3}{2\pi},
\beta_1=1+\frac{i\theta_2}{2\pi},
\beta_2=  \frac12(1+\rho)+\frac{i\theta_3}{2\pi}$
for the two cases $\rho=\pm1$. We have numerically checked that
the estimate (\ref{F32as}) is indeed valid for these cases in the range 
$|\theta_{23}|<4\pi$. It is plausible that it can also be proved for
arbitrary values of $\Re\,\theta_{23}$ for some range of $\Im\,\theta_{23}$
by assuming analyticity of the formula in this variable.
Applying then (\ref{F32as}) to the first terms of (\ref{f1ex}-\ref{f3ex}) 
(for $s=-1/4$) we find
\begin{equation}
f^{(I)}_m(\theta_1,\theta_2,\theta_3)\approx\frac{ic}{2}\,
{\rm e}^{-\frac{1}{2}\theta_1}\,
\frac{{\rm e}^{\frac{1}{2}(\theta_2+\theta_3)}}
{\theta_{12}^{3/2}}\,\left(\begin{matrix}-1\\1\\2\end{matrix}\right)\,.
\end{equation}
Thus the first terms only contribute to $W_-$ for $s=-1/4$. Their 
contribution is:
\begin{equation}
\left[W^{(-)}_-(\xi)\right]^{(I)}=-ic\,{\rm e}^{-\frac{1}{2}\xi}\,.
\end{equation}

For the second and third terms (of $f_1$ and $f_2$) we have to
consider, for large positive $\lambda=\frac{\theta_{12}}{2\pi}$,
the following product:
\begin{equation}
\begin{split}
\Gamma(a_1)\Gamma(a_2)\Gamma(-a_1-a_2)&\Gamma(\alpha_3+i\lambda)
\Gamma\left(\frac{1}{2}-\alpha_3-i\lambda\right)\\
&_3F_2\left(a_1,a_2,\alpha_3+i\lambda;
1+a_1+a_2,\frac{1}{2}+\alpha_3+i\lambda;1\right)\,.
\end{split}
\label{prod}
\end{equation}
Its asymptotic form can be established using the integral
representation (\ref{int32}) together with the formula \cite{AbSt}
\begin{equation}
\begin{split}
_2F_1&(a_1,a_2;1+a_1+a_2;t)\approx \frac{\Gamma(1+a_1+a_2)}
{\Gamma(a_1)\Gamma(a_2)}\\
&\left\{\frac{1}{a_1a_2}+(1-t)\left[
\ln(1-t)-\Psi(1)-\Psi(2)+\Psi(1+a_1)+\Psi(1+a_2)\right]\right\}\,,
\end{split}
\end{equation}
valid in the vicinity of $t=1$. With the help of this formula
we can calculate the large $\lambda$ expansion of (\ref{prod}):
\begin{equation}
\frac{2\pi^2i}{\sqrt{\lambda}}\,{\rm e}^{\frac{i\pi}{4}}\,
\frac{{\rm e}^{-\pi\lambda}\,{\rm e}^{i\pi\alpha_3}}
{\sin\pi(a_1+a_2)}\,\left\{\frac{1}{a_1a_2}-\frac{i}{2\lambda}
\left[X-\ln\lambda-\frac{i\pi}{2}+\Psi\left(\frac{3}{2}\right)
\right]\right\}+\cdots,
\label{as23}
\end{equation}
where
\begin{equation}
X=\Psi(1+a_1)+\Psi(1+a_2)-\Psi(1)-\Psi(2)+\frac{\frac{1}{4}-\alpha_3}
{a_1a_2}\,.
\end{equation}
Using (\ref{as23}) in (\ref{f1ex}) and (\ref{f2ex}) we find the 
following results. For $s=1/4$ the 
$\frac{{\rm e}^{-\frac{1}{2}\theta_{12}}}{\theta_{12}^{1/2}}$ terms
cancel and only the
$\frac{{\rm e}^{-\frac{1}{2}\theta_{12}}}{\theta_{12}^{3/2}}$ terms
remain. In this case these contribute to the leading terms
$U_m$ and we find
\begin{equation}
U^{(+)}(\xi)=\frac{ic}{2\cosh\frac{\xi}{2}}\,,
\end{equation}
the same result as we found in the previous subsection. For $s=-1/4$
(\ref{as23}) gives contributions both to the leading $U_m$ and the 
subleading $W_m$ terms. We find
\begin{equation}
U^{(-)}(\xi)=\frac{2c}{\pi}\,\frac{{\rm e}^{-\frac{1}{2}\xi}}
{\xi-i\pi}\,,
\end{equation}
\begin{equation}
W_+^{(-)}(\xi)=-\frac{c}{\pi}\,{\rm e}^{-\frac{1}{2}\xi}\,
\frac{\xi-2\pi i}{\xi-i\pi}
\end{equation}
and
\begin{equation}
\begin{split}
\left[W^{(-)}_-(\xi)\right]^{(II)+(III)}=
&-\frac{2c}{\pi}\,{\rm e}^{-\frac{1}{2}\xi}\,
\Big[
\frac{1}{4}\Psi\left(\frac{1}{2}+\frac{i\xi}{2\pi}\right)+
\frac{1}{4}\Psi\left(\frac{1}{2}-\frac{i\xi}{2\pi}\right)\\
&+\frac{1}{2}\Psi\left(\frac{3}{2}\right)
+\frac{1}{2}\Psi\left(\frac{1}{2}\right)-\Psi(1)
-\frac{1}{2}\ln\frac{\Delta}{2\pi}-\frac{i\pi}{2}\Big]\,.
\end{split}
\end{equation}
Again, the above results are in agreement with the ones obtained in
the previous subsection solving the asymptotic form factor equations. 
Finally we get
\begin{equation}
g_0=\Psi\left(\frac{1}{2}\right)-\Psi(1)+1-
\frac{1}{2}\ln\frac{\Delta}{2\pi}=1-
\frac{1}{2}\ln\frac{8\Delta}{\pi}\,.
\end{equation}

% List of references

\eject

\end{document}